\documentclass[pdflatex,sn-mathphys-num]{sn-jnl}% Math and Physical Sciences Numbered Reference Style
\usepackage{color,morefloats}
\usepackage{latexsym, amsthm}
\usepackage{setspace}
\usepackage[utf8]{inputenc}
\usepackage[T1]{fontenc}
\usepackage{siunitx}
\usepackage{colortbl}
\usepackage{caption}
\usepackage{float}
\usepackage{titlesec}
\usepackage{mathtools}

\usepackage{graphicx}%
\usepackage{multirow}%
\usepackage{amsmath,amssymb,amsfonts}%
\usepackage{amsthm}%
\usepackage{mathrsfs}%
\usepackage[title]{appendix}%
\usepackage{xcolor}%
\usepackage{textcomp}%
\usepackage{manyfoot}%
\usepackage{booktabs}%
\usepackage{algorithm}%
\usepackage{algorithmicx}%
\usepackage{algpseudocode}%
\usepackage{listings}%
\theoremstyle{thmstyleone}%
\newtheorem{theorem}{Theorem}%  meant for continuous numbers
\newtheorem{proposition}[theorem]{Proposition}% 

\theoremstyle{thmstyletwo}%
\newtheorem{example}{Example}%
\newtheorem{remark}{Remark}%
\newtheorem{observation}{Observation}

\theoremstyle{thmstylethree}%
\newtheorem{definition}{Definition}%

\newcommand{\F}{\mathbb F}
\newcommand{\B}{\mathbb B}

\newcommand{\Tr}{\operatorname{Tr}_{\F/\B}}

\newcommand{\spanB}{\operatorname{span}_\B}

\newcommand{\rankB}{\operatorname{rk}_\B}
\newcommand{\dimB}{\operatorname{dim}_\B}

\newcommand{\im}{\operatorname{Im}}

\newcommand{\Per}{\operatorname{Per}}

\newcommand{\GRS}{\operatorname{GRS}}
\newcommand{\RS}{\operatorname{RS}}
\newcommand{\M}{\mu}
\newcommand{\GRM}{\operatorname{GRM}}

\newcommand{\TB}{\operatorname{TB}}

\newcommand{\dTB}{d_{loc}}

\newcommand{\Cbase}{C_0}
\newcommand{\Gbase}{G_0}

\newcommand{\nbase}{n_0}
\newcommand{\kbase}{k_0}
\newcommand{\Dbase}{D_0}

\newcommand{\greyMidrule}{%
\arrayrulecolor{gray!50}\midrule\arrayrulecolor{black}%
}

\begin{document}

\title[Systematic Derivation of Leakage in Massey Secret Sharing Schemes]{Systematic Derivation of Leakage in Massey Secret Sharing Schemes}

%%=============================================================%%
%% GivenName	-> \fnm{Joergen W.}
%% Particle	-> \spfx{van der} -> surname prefix
%% FamilyName	-> \sur{Ploeg}
%% Suffix	-> \sfx{IV}
%% \author*[1,2]{\fnm{Joergen W.} \spfx{van der} \sur{Ploeg} 
%%  \sfx{IV}}\email{iauthor@gmail.com}
%%=============================================================%%

\author[1,2,3]{\fnm{Nadja} \sur{Aoutouf}}\email{nadja.aoutouf@inria.fr}
\equalcont{These authors contributed equally to this work.}

\author[1,2,3]{\fnm{Daniel} \sur{Augot}}\email{daniel.augot@inria.fr}
\equalcont{These authors contributed equally to this work.}

\affil[1]{\orgname{INRIA}}
\affil[2]{\orgname{École Polytechnique}}
\affil[3]{\orgname{LIX, CNRS UMR 7161}}

%\affil[3]{\orgdiv{Department}, \orgname{Organization}, \orgaddress{\street{Street}, \city{City}, \postcode{610101}, \state{State}, \country{Country}}}

%%==================================%%
%% Sample for unstructured abstract %%
%%==================================%%

\abstract{
    Leakage attacks on secret sharing schemes can be based on the exploitation of partial information about 
    individual shares to recover the underlying secret. In coding theory, 
    linear exact repair schemes (LERSs) allow the recovery of a codeword symbol 
    from limited information obtained from the remaining symbols. 
    When applied to code-based secret sharing schemes, such repair schemes can be interpreted as 
    recovering the secret from partial information about the shares, 
    expressed in terms of subfield symbols. Guruswami and Wootters~\cite{GW16} 
    introduced the first explicit constructions of
    LERSs for low-rate Reed--Solomon codes that is optimal in terms of bandwidth.

    In this work, we first investigate the construction and algebraic properties of LERSs, 
    building on the Guruswami--Wootters construction. 
    We derive new explicit repair schemes from their construction and establish 
    structural properties of the resulting schemes.

    Second, we further develop a systematic computational approach based on a Gröbner 
    bases approach to identify leakage functions associated with LERSs for arbitrary 
    linear codes. This approach enables us to investigate the 
    existence and structural properties of LERSs across different code families 
    and parameter ranges. We first applied our method to Reed--Solomon codes, 
    which gave as further insights into the algebraic structure of LERSs. 
    Moreover, by applying our method to generalized Reed--Muller 
    and Tamo--Barg codes, we identify several LERSs that, 
    to the best of our knowledge, have not previously been described in the literature.

}


%%\pacs[JEL Classification]{D8, H51}

%%\pacs[MSC Classification]{35A01, 65L10, 65L12, 65L20, 65L70}

\maketitle

\section{Introduction}
Secret sharing schemes are a fundamental building block of modern cryptography. 
Their security is well understood in the traditional setting, where an adversary 
obtains complete shares from a subset of participants. However, considerably less 
is known about their security when an adversary can simultaneously obtain partial information 
from multiple shares. This setting is motivated by side-channel attacks, 
particularly probing attacks, in which an adversary observes limited information from different 
shares. Motivated by this concern, a line of research initiated by~\cite{BDIR19} and 
culminating in~\cite{TCC:Kasser24,EC:Nguyen25,BS26} has investigated the leakage resilience of 
secret sharing scheme, with particular emphasis on determining its leakage-resilience threshold.

Linear codes play an important role in the construction of secret sharing schemes. 
In particular, \emph{Massey secret sharing schemes} are constructed from linear codes, 
where
the relations among code symbol determine how 
information can be reconstructed from shares and how additional linear observations may reveal 
information about the secret. Understanding these relations is therefore important for analyzing 
the leakage resilience of Massey secret sharing schemes.
\emph{Linear exact repair schemes} (LERS) provide a framework for studying such relations. 
They characterize how a symbol of a codeword can be reconstructed from linear 
information obtained from the remaining symbols. 
Beyond their role in distributed storage, these schemes are closely connected to the structure of 
the dual code and can be used to investigate leakage in the corresponding Massey secret sharing schemes. 
However, obtaining explicit repair schemes for a given code, and understanding 
the bandwidth they require, can be challenging.

Guruswami and Wootters~\cite{GW16} give a first 
construction of a LERS
for Reed--Solomon codes,
which directly yields leakage functions for Shamir's secret sharing scheme. Moreover, their
provide a general characterization of LERS for any Massey secret sharing scheme. 
While the characterization of Guruswami and Wootters provides a theoretical framework for linear exact repair, 
finding explicit repair schemes for a given code remains a challenging problem. 
In particular, existing constructions do not provide a systematic way to derive all existing LERSs with small bandwidth across different code families.
This raises the following questions: How can we computationally identify linear exact repair schemes with bandwidth at most $n-1$ for a given 
linear code? What can the resulting schemes reveal about the repair properties of different codes?
Motivated by these questions, we develop a computational approach based on the general characterization of Guruswami and 
Wootters~\cite{GW16}. Our approach allows us to search for explicit repair schemes across different linear codes, 
including Reed--Solomon, generalized Reed--Muller, and Tamo--Barg codes. 
Due to the direct correspondence in-between LERSs and leakage functions, the resulting schemes also provide explicit leakage functions for the 
corresponding Massey secret-sharing schemes.

\subsection{Our Contribution}
The main contributions of this work are as follows.
First, we study algebraic constructions and structural properties of the explicit 
LERSs by Guruswami and Wootters:
\begin{itemize}
	\item \textbf{Derivation of new LERSs.} We investigate several ways in which the Guruswami--Wootters 
	construction can give rise to further repair schemes. In particular, we consider Reed--Solomon codes of 
	lower dimension, 
	introduce a reverse transformation of check-polynomials that produces new LERSs, 
	and study the invariance of the construction under suitable permutations of the evaluation points. 
	These results reveal additional structure in the family of LERSs obtained from the 
	Guruswami--Wootters construction.
\end{itemize}
Furthermore, for an arbitrary linear code $C\subseteq\mathbf{F}_{q^m}^n$, we design and implement a systematic 
method for deriving all 
uniform LERSs of bandwidth at most $n-1$, and, equivalently, 
the corresponding leakage functions:
\begin{itemize}
	\item \textbf{Development of a Systematic Method.}
	We develop a systematic method for deriving all uniform LERSs 
	of bandwidth at most $n-1$ for an arbitrary linear code $C$. In particular, we identify a polynomial 
	system whose solutions are equivalent to the existence 
	of a LERS and employ computational algebra techniques as Gröbner basis computations to determine all solutions.
	\item \textbf{Efficiency Improvements.}
	To improve the efficiency and scalability of the systematic method, we include a hybrid computational 
	approach that combines Gröbner basis computations with variable fixing, resultant-based elimination, and distributed computation.
	\item \textbf{Experimental Evaluation.}
	We apply the resulting program to Reed--Solomon, Reed--Muller and Tamo--Barg codes for a range of parameters, 
	including parameter regimes that, to the best of our knowledge, have not been considered previously. 
	\item \textbf{Algebraic Properties of LERSs.}
	Motivated by the experimental observations, we formulate and prove further algebraic properties of LERSs. 
\end{itemize}

\section{Preliminaries}
Throughout this work, for a positive integer $n$, we use the notation
$[n]_0 := \{0,1,\ldots,n\}$
 and 
$[n] := \{1,\ldots,n\}.$
For a prime power q, we denote by $\B=\mathbf{F}_q$ the finite field of size $q$ and by
$\F$ an extension field of $\B$ of degree $m$. We also define $\F[X]$ to be the set of polynomials over $\F$
and denote by $\Tr$ the field trace map from $\F$ to $\B.$ 
Let $\zeta=(\zeta_1,\ldots,\zeta_m)$ be a basis of $\F$ over $\B$.
For a matrix $M\in\F^{u\times v}$, we denote by
$M_\B\in\B^{um\times v}$ the matrix obtained by expanding each
entry of $M$ with respect to $\zeta$, column-wise. More precisely,
the rows of $M_\B$ are indexed by pairs $(i,\ell)$, and
$(M_\B)_{(i,\ell),j}=(M_{ij})_\ell,$
where $(M_{ij})_\ell$ denotes the $\ell$-th coordinate of $M_{ij}$
with respect to $\zeta$. Similarly, we denote by $M^\B\in\B^{u\times
vm}$ the row-wise expansion of $M$.
\subsection{Linear Codes}
%%%%%%%%%%%%%%%%%%%%%%%
% Basics of Linear Codes
%%%%%%%%%%%%%%%%%%%%%%%
The following definitions and results closely follow~\cite{AC20} and~\cite{R6}. 
We begin by recalling some basic definitions concerning linear codes.
A $[n,k]_{\F}$ linear code $C$ is a $k$-dimensional subspace of $\F^n$.
A generator matrix $G\in\F^{k\times n}$ has rows spanning $C$ (i.e.: $C=\left\{xG:x\in\F^k\right\}$), while
a parity-check matrix $H\in\F^{(n-k)\times n}$ has rows spanning its dual
$C^\perp:=\{h\in\F^n:h\cdot c=0\text{ for all }c\in C\}.$
In particular, $GH^{\mathsf T}=\mathbf{0}$.

For $x\in\F^n$, its Hamming weight $\operatorname{w}_H(x)$ is the number
of non-zero entries (i.e.: $\operatorname{w}_H(x) := \left|\left\{i\in[n-1]_0:x_i\neq 0\right\}\right|$). 
The Hamming distance is
$d_H(x,y):=\operatorname{w}_H(x-y)$, and the minimum Hamming distance
of a non-zero linear code $C$ is
\[
d_H(C):=\min_{x\in C\setminus\{\mathbf{0}\}}\operatorname{w}_H(x).
\]
The Singleton bound states that $d_H(C)\leq n-k+1$.
A $[n,k]_{\F}$ linear code $C$ is called \emph{maximum-distance
separable} (MDS) if it attains the Singleton bound.

\subsubsection{Explicit Codes}
We introduce several families of evaluation-based codes that will be used throughout this work. 
%%%%%%%%%%%%%%%%%%%%%%%
% RS-Code and GRS-Code
%%%%%%%%%%%%%%%%%%%%%%%

Let $1\leq k\leq n\leq |\F|$, let
$\alpha=(\alpha_0,\ldots,\alpha_{n-1})\in\F^n$ consist of distinct
elements, and let $\gamma=(\gamma_0,\ldots,\gamma_{n-1})
\in(\F\setminus\{0\})^n$.
The \emph{generalized Reed--Solomon} code is defined as
\[
\GRS(\mathbb{\alpha},\mathbb{\gamma},k)
:=
\left\{
\bigl(\gamma_0f(\alpha_0),\ldots,\gamma_{n-1}f(\alpha_{n-1})\bigr)
\;\middle|\;
f\in\F[X],\ \deg(f)\leq k-1
\right\}.
\]
The \emph{Reed--Solomon} code is the special case
$\RS(\alpha,k):=\GRS(\alpha,\mathbf{1},k)$.
Note that, both are $[n,k]_\F$ MDS codes.
The dual of a generalized Reed--Solomon code is again a generalized Reed--Solomon code:
\[
\GRS(\mathbb{\alpha},\mathbb{\gamma},k)^\perp
=
\GRS(\mathbb{\alpha},\mathbb{\gamma}',n-k),
\]
where $\mathbb{\gamma}'=(\gamma'_0,\ldots,\gamma'_{n-1})$ is defined by
$(\gamma'_i)^{-1}:=\gamma_i\prod_{\substack{j\in[n-1]_0\backslash\{i\}}}(\alpha_i-\alpha_j).$
In particular, if $n=|\F|$, then
$\RS(\alpha,k)^\perp=\RS(\alpha,n-k).$
Thus, every dual codeword $c^\perp\in \GRS(\alpha,\gamma,k)^\perp$ 
can be described by 
an associated polynomial $h\in\F[X]$ with $\deg(h)\leq n-k-1$ such that
$c^\perp =\bigl(\gamma'_0 h(\alpha_0),\ldots,\gamma'_{n-1}h(\alpha_{n-1})\bigr),$
where $h\in\F[X]$ is called the \emph{check polynomial}
associated with $c^\perp$.

%%%%%%%%%
% RM-Code
%%%%%%%%%

We next introduce generalized Reed--Muller (GRM) codes (first published in~\cite{KLP68}), 
which extend RS codes to multivariate
polynomials. Let $\alpha=(\alpha_0,\ldots,\alpha_{n-1})$ enumerate
$\F^\M$, where $n=(q^m)^\M$. The GRM code of order $\rho$ in $\M$
variables\footnote{Note that, for Reed--Muller codes, 
 we use the Greek letters $\rho$ and $\M$ for the parameters, 
 since the standard notation $\GRM(r,m)$ with
 $r$ and $m$ is already reserved for other parameters.} is defined as
    \[
    \GRM(\rho,\M):=\left\{\bigl(f(\alpha_0),\ldots,f(\alpha_{n-1})\bigr)\;\middle|\;f\in\F[X_1,\ldots,X_\M],\ \deg(f)\leq\rho\right\}.
    \]
For $\M=1$ and $\rho<q^m$, this reduces to a full-length RS code.
Let $0\leq\rho<\M(q^m-1)$ and write
$\rho=\gamma(q^m-1)+\theta$, where $0\leq\theta<q^m-1$.
The minimum Hamming distance is~\cite{DGM70}:
$d_H(\GRM(\rho,\M))
=(q^m-\theta)(q^m)^{\M-\gamma-1}.$
Moreover, its dual is again a GRM code, with
$$\GRM(\rho,\M)^\perp
=\GRM(\M(q^m-1)-\rho-1,\M),$$
and minimum Hamming distance~\cite{DGW+10} is given by
$d_H(\GRM(\rho,\M)^\perp)=(\theta+2)(q^m)^\gamma.$

%%%%%%%%%
% TB-Code
%%%%%%%%%
Evaluation codes can also be designed to provide additional properties, such as locality.
A linear code has \emph{locality} $r$ if every coordinate can be
recovered from at most $r$ other coordinates, called its
\emph{recovery set}. Tamo--Barg (TB) codes~\cite{TB13} are
evaluation-based locally recoverable codes constructed as follows.
Let $(r+1)\mid n$ and let $A\subseteq\F$ consist of $n$ distinct
elements partitioned into \emph{locality sets}
\[
A=A_1\sqcup\cdots\sqcup A_{n/(r+1)},\qquad |A_\ell|=r+1.
\]
Let $g\in\F[X]$ be a polynomial of degree $r+1$ that is constant
on each $A_\ell$. Writing $k=ur+v$, with $1\leq v\leq r$, define
\[
\mathcal P:=
\operatorname{span}_{\F}
\left(
\{X^i g(X)^j:0\leq i<r,\ 0\leq j<u\}
\cup
\{X^i g(X)^u:0\leq i<v\}
\right).
\]
The \emph{Tamo--Barg code} is defined as
\[
\TB(n,k,r)
:=
\left\{
\bigl(f(\alpha_0),\ldots,f(\alpha_{n-1})\bigr)
\;\middle|\;
f\in\mathcal{P}
\right\}.
\]
Note that, for $r=k$, the construction reduces to a Reed--Solomon code.
\begin{remark}
	The Tamo--Barg construction can be defined more generally with a \emph{local distance} 
	parameter $\dTB \geq2$, in which case each locality set has size $r+\dTB-1$ 
	and the restriction of the code to each locality set is a $[r+\dTB-1,r]_\F$ 
	Reed--Solomon code. However, our work focuses on the case $\dTB=2$.
\end{remark}
\begin{example}[Standard construction of $g$ using multiplicative cosets]\label{ex:tb:good_poly_construction}
	Let $(r+1)\mid(q^m-1)$ and $H\leq\F\backslash\{0\}$ be a subgroup of order $r+1.$ 
    Define 
	\begin{equation}\label{eq:tb:good_poly}
		g(X):=\prod_{h\in H}(X-h)=X^{r+1}-1,
	\end{equation}
	where the equality follows from our choice of
	 $H$.
	 Note that such a subgroup 
	exists since $(r+1) \mid (q^m-1).$
	For any $a\in\F\backslash\{0\}$ we have $aH=\{ah\colon h\in H\}.$
	Hence, for some $\alpha\in aH$, there exist $h\in H$ such that $\alpha=a h,$
	which implies $\alpha^{r+1}=a^{r+1}$.
	Thus,
	$g(\alpha)=a^{r+1}-1.$
	Consequently, $g$ is constant on every multiplicative coset $aH$ of $H$.
	Therefore, we can choose the locality sets as distinct cosets of $H$ in $\F\backslash\{0\}$.
\end{example}
%%%%%%%%%
% Cyclic-Code
%%%%%%%%%

A linear code $C\subseteq\F^n$ is called \emph{cyclic}~\cite{HP10}
if for every
$c=(c_0,\ldots,c_{n-1})\in C$ we have $(c_1,\ldots,c_{n-1},c_0)\in C$.
The permutation group of $C$ is
$\Per(C):=\{\pi\in S_n:\pi(C)=C\}.$
It is known that for every linear code $C$, we have
$\Per(C)=\Per(C^\perp).$
In particular, let $a\in\F\backslash\{0\}$ be a generator of $\F\backslash\{0\}$
and $\alpha=(1,a,\ldots,a^{q^m-2})$. Then, for every
$1\leq k\leq q^m-1$, the Reed--Solomon code
$\RS(\alpha,k)$ is cyclic~\cite{KLP68}.

\subsubsection{Shortening and Puncturing}\label{Ch2::ShorteningPuncturing}
Let $C$ be a $[n+1,k]_{\F}$ code and let $I\subseteq [n]_0$ be an
index set. The \emph{puncturing} of $C$ at $I$, denoted by
$\underline{C}_I$, is the code of length $n+1-|I|$ obtained by
deleting the coordinates indexed by $I$:
\[
\underline{C}_I
:=
\left\{
(c_j)_{j\in [n]_0\setminus I}
\;\middle|\;
(c_0,\ldots,c_n)\in C
\right\}.
\]
The \emph{shortening} of $C$ at $I$, denoted by $\overline{C}_I$,
is the code obtained by restricting to codewords that vanish on the
coordinates in $I$ and subsequently deleting these coordinates:
\[
\overline{C}_I
:=
\left\{
(c_j)_{j\in [n]_0\setminus I}
\;\middle|\;
(c_0,\ldots,c_n)\in C,\;
c_i=0\text{ for all }i\in I
\right\}.
\]
If $H$ is a parity-check matrix of $C$, then a parity-check matrix of
$\overline{C}_I$ is obtained by deleting the columns indexed by $I$
from $H$. Consequently,
$\overline{(C^\perp)}_I=(\underline{C}_I)^\perp.$
When the context is clear, we omit the index set $I$ and simply write
$\underline{C}$ and $\overline{C}$.

\subsection{Massey Secret Sharing Schemes}
We use the well-established connection between linear codes and secret sharing 
schemes~\cite{massey-secret-sharing93}, and follow the terminology 
of~\cite{doron2026discrepancyrandomlinearcodes}. For this purpose, we consider 
a $[n+1,k]_{\F}$ code $C$ that admits a generator 
matrix $G$ of the form 
\begin{equation} 
	G= \begin{bmatrix} -1 & \phi\\ 
		0 & \overline{G} \end{bmatrix}, 
		\label{eq:G:non:degenerate} 
\end{equation} 
where $\phi\in\F^{n}$ is nonzero and $\overline{G}$ is a generator matrix of the shortening 
of $C$ at the first coordinate. Let $s\in\F$ be a secret. 
Then the dealer samples a codeword 
$c=(c_0,\ldots,c_{n})\in C$
with the constraint $c_0=s$. 
The value $c_0$ represents the secret, while $c_1,\ldots,c_{n}$ are the corresponding shares. 
Let $d$ and $d^\perp$ denote the minimum Hamming distances of this code
$C$ and 
the dual code $C^\perp$, respectively. 
Then any set of $d^\perp-2$ shares provides no information about the secret, 
whereas any set of $n-d+2$ shares allow reconstructing it.

% Note that, every $\underline{h}\in\underline{\Dbase}$ admits a
% unique extension $(h_0,\underline{h})\in\Dbase$, where
% $h_0=\phi\cdot\underline{h}^{T}.$
% We refer to the $\B$-linear map
% 	\begin{equation}\label{eq:phi:D}
% 			\phi_{\Dbase} \colon \left\{\begin{aligned}
% 				\underline{\Dbase}& \longrightarrow \F \\
% 				\underline{h}& \longmapsto \phi\cdot\underline{h}^{T}
% 			\end{aligned} \right.
%     \end{equation}
% as the \emph{extension map} associated with $\Dbase$.
\begin{example}[Shamir's secret sharing scheme]
Shamir's secret sharing scheme~\cite{S79} can be viewed as a Massey secret sharing scheme based on a Reed--Solomon code.
To share a secret $s\in\F$, let $\alpha_0,\alpha_1,\ldots,\alpha_{n}\in\F$ be 
pairwise distinct evaluation points with $\alpha_0=0$,
and choose a polynomial
$f\in\F[X]$
uniformly at random among all polynomials of degree less than $k$ satisfying
$f(0)=s.$
The secret is then $s=f(\alpha_0)$, while the $n$ shares are given by
$f(\alpha_1),\ldots,f(\alpha_{n}).$
Thus, $C$ is a $[n+1,k]_{\F}$ Reed--Solomon code given by
\[
    C
    =
    \left\{
    \bigl(f(\alpha_0),f(\alpha_1),\ldots,f(\alpha_{n})\bigr)
    \,\middle|\,
    f\in\F[X],\ \deg(f)<k
    \right\},
\]
and the generator matrix of $C$ can be written as
\[
    G=
    \begin{bmatrix}
        -1 & -1 & \cdots & -1\\
        0 & \alpha_1 & \cdots & \alpha_{n}\\
        0 & \alpha_1^2 & \cdots & \alpha_{n}^2\\
        \vdots & \vdots & & \vdots\\
        0 & \alpha_1^{k-1} & \cdots & \alpha_{n}^{k-1}
    \end{bmatrix}.
\]
Hence, in the notation above, we have
   $ \phi=(-1,\ldots,-1)\in\F^{n}.$
\end{example}

\subsection{Explicit LERSs including the Guruswami--Wootters Construction}\label{Ch2::LERS:GW}
\renewcommand{\nbase}{n}
\renewcommand{\kbase}{k}
\renewcommand{\Cbase}{C}
\renewcommand{\Gbase}{G}
\renewcommand{\Dbase}{D}
The reconstruction of the secret in a Massey secret sharing scheme can
be viewed as a linear exact repair problem: Given a $[n+1,k]_\F$ code $C$ and a codeword $c\in C$, 
then the secret coordinate
$c_0$ is recovered from information obtained from the remaining code
symbols $c_1,\ldots,c_n$. 
Following~\cite{GW16}, we give a simplified definition of a LERS.
\begin{definition}\label{def:LERS}
	A uniform \emph{linear exact repair scheme} (LERS) for a
	$[n+1,k]_{\F}$ code $C$ with respect to a subfield
	$\B\subseteq\F$ and an integer $s \leq m$ consists of
	\begin{itemize}
		\item $\B$-linear maps $g_j\colon\F\rightarrow\B^s$ for
		$j\in[n]$,
		\item a $\B$-linear reconstruction map
		$R\colon\B^{sn}\rightarrow\F$,
	\end{itemize}
	such that, for every $c=(c_0,\ldots,c_n)\in C$, we have
	$c_0=R\bigl(g_1(c_1),\ldots,g_n(c_n)\bigr).$
\end{definition}
The \emph{repair bandwidth} is the total number of downloaded
$\B$-symbols, given by
\[
b:=\sum_{j\in[n]}\dim_\B(\operatorname{Im}(g_j)).
\]
Throughout this work, we restrict our attention to the case $s=1$, and hence $b\leq n$.

The main result of Guruswami and Wootters shows that 
Reed--Solomon codes over $\F$ admit a LERS over $\B$ when their
dimension is sufficiently small.
\begin{theorem}[\cite{GW16}]\label{gw:rs:LERS_construction}
	Let $[\F:\B]=m$, $n+1\leq q^m$, and
	$n+1-k\geq q^{m-1}$. Then the Reed--Solomon code
	$\RS(\alpha,k)$ of length $n+1$
	admits a LERS over $\B$.
\end{theorem}

More generally, Guruswami and Wootters characterize the existence of
such a repair scheme of a linear code in terms of suitable dual codewords.
\begin{theorem}[\cite{GW16}]\label{thm:gw:LERSexistince}
	Let $\Cbase$ be a base code $[\nbase+1,\kbase]_{\F}$ code. If there
	exist $m$ dual codewords
	$h^{(1)},\ldots,h^{(m)}\in C^\perp$ satisfying
	\begin{align}
		\dim_\B\!\left(
			\operatorname{span}_\B
			\{h_0^{(1)},\ldots,h_0^{(m)}\}
		\right)
		&=m,
		\label{eq:gw:basic:m}\\
		\dim_\B\!\left(
			\operatorname{span}_\B
			\{h_j^{(1)},\ldots,h_j^{(m)}\}
		\right)
		&\leq 1,
		\qquad \forall\,j\in[\nbase],
		\label{eq:gw:basic:1}
	\end{align}
	then there exists a LERS requiring at most $\nbase$ $\B$-symbols to
	repair the erased symbol $c_0$.
\end{theorem}

A LERS induces a leakage function describing the
information downloaded from the remaining shares.
\begin{definition}[Leakage function]
Let $\bigl((g_j)_{j\in[n]},R\bigr)$ be a LERS for a code
$C\subseteq\F^{n+1}$. Its associated \emph{leakage function} is
the $\B$-linear map
\begin{equation*}
    \mathsf{Leak}\colon\left\{\begin{aligned}\F^n &\longrightarrow \B^n\\
        (c_1,\ldots,c_n) &\longmapsto
        \bigl(g_1(c_1),\ldots,g_n(c_n)\bigr).
    \end{aligned}
    \right.
\end{equation*}
By the definition of a LERS, every $c\in C$ satisfies
$c_0=R\bigl(\mathsf{Leak}(c_1,\ldots,c_n)\bigr).$
\end{definition}

We next examine known LERS constructions and their associated
leakage functions.

\subsubsection{Reed-Solomon Codes}\label{subsec:LERS_RScode}
Consider a Reed--Solomon code $\RS(\alpha,k)_\F$ of length $n+1$. Guruswami and Wootters~\cite{GW16}
construct a LERS for recovering $c_0=f(\alpha_0)$ using the $m$ check polynomials
\begin{equation}\label{eq:gw:check-poly}
h_i(X)=\frac{\Tr(\zeta_i(X-\alpha_0))}{X-\alpha_0},
\qquad i\in[m],
\end{equation}
where $\zeta_1,\ldots,\zeta_m$ form a $\B$-basis of $\F$.
Since $h_i(\alpha_0)=\zeta_i$ and
$h_i(\alpha_j)=\Tr\bigl(\zeta_i(\alpha_j-\alpha_0)\bigr)/(\alpha_j-\alpha_0),$ for $j\in[n],$
the check polynomials satisfy conditions~\eqref{eq:gw:basic:m}
and~\eqref{eq:gw:basic:1}. The resulting leakage function is
$\mathsf{Leak}(c_1,\ldots,c_n)=(g_1(c_1),\ldots,g_n(c_n)),$ where
\begin{equation}\label{eq:gw:leakage_fct}
g_j:c_j\longmapsto
\Tr\left(\frac{\gamma_j c_j}{\alpha_j-\alpha_0}\right),
\qquad j\in[n].
\end{equation}
The construction applies whenever $n+1\leq q^m$ and
$n+1-k\geq q^{m-1}$.

Dau and Milenkovic~\cite{DM17} generalized this approach using
subspace polynomials. For an $s$-dimensional $\B$-subspace
$W\subseteq\F$, let
$L_W(X):=\prod_{w\in W}(X-w).$
The corresponding check polynomials are
\begin{equation}\label{eq:dm:check-poly}
h_i(X)=\frac{L_W(\zeta_i(X-\alpha_0))}{X-\alpha_0},
\qquad i\in[m],
\end{equation}
which satisfy~\eqref{eq:gw:basic:m} and the weaker dimension bound
$\dimB\left(\spanB\bigl(h_1(\alpha_j),\ldots,h_m(\alpha_j)\bigr)\right)\leq m-s,$
for $j\in[n].$
Consequently, each node (a.k.a. share) $i$ for which $\alpha_i\notin W$ contributes at most $m-s$
$\B$-symbols.
The construction applies to Reed--Solomon codes of length
$n+1\leq q^m$, provided that $n+1-k\geq q^s$.

In~\cite{DGW+10} it is shown, that for a $[n+1,k]_\F$ MDS code, repairing one symbol from $d$
helper nodes requires a repair bandwidth of at least
\begin{equation}\label{eq:bandwidth:cut-set-bound}
b\geq\frac{dm}{d-k+1}.
\end{equation}
In~\cite{TYB17} an explicit construction of a LERS for Reed--Solomon codes over a field $\F$ that
attains this cut-set bound is given, provided that the extension degree $m$ is chosen large enough
(i.e.; $m\approx n^n$). Note that due to the required size of the extension degree, this construction is rather theoretical and impractical. 
Nevertheless, we recall their construction and
make the associated leakage function explicit.
Let $p_0,\ldots,p_{\ell-1}$ denote the primes not exceeding
$n+1-k$, where $\ell$ is the number of such primes. Choose
$m=\prod_{i=0}^{\ell-1}p_i,
\F=\mathbf{F}_{q^m}, q\geq n+1-\ell.$
For each $i\in[\ell-1]_0$, choose $\alpha_i\in\F$ such that
$[\B(\alpha_i):\B]=p_i$, and define
$\F_i:=\B(\alpha_j:j\neq i).$
Then $[\F:\F_i]=p_i$. The remaining $n+1-\ell$
evaluation points are chosen as distinct elements of $\B$.
Consider the resulting code $\RS(\alpha,k)$.
Then the symbol on position $i$ can be repaired from any $d_i=p_i+k-1$
symbols. We focus on repairing $c_0$ from an arbitrary
set $R_0\subseteq[n]$ of size $d_0$.
Define the annihilator polynomial
$
a_0(X)
=
\prod_{j\in[n]_0\setminus(R_0\cup\{0\})}
(X-\alpha_j).
$
The construction uses the following $p_0$ check polynomials:
$$
h_{0,s}(X)=X^s a_0(X),
\qquad
s\in[p_0-1]_0.
$$
Since $\deg(h_{0,s})\leq n-k$, these polynomials induce
valid parity-check equations.
Applying these equations and the trace map
$\operatorname{Tr}_{\F/\F_0}$
yields the leakage function
\begin{equation}\label{eq:primetower:leakagefct}
	g_j:c_j\longmapsto
	\operatorname{Tr}_{\F/\F_0}
	\left(
	\gamma_j a_0(\alpha_j)c_j
	\right),
	\qquad
	j\in R_0,
\end{equation}
where $\gamma_j$ are the dual Reed--Solomon multipliers.
These values determine the $p_0$ traces
$\operatorname{Tr}_{\F/\F_0}\left(
\gamma_0a_0(\alpha_0)\alpha_0^s c_0
\right),$
for all $s\in[p_0-1]_0.$
Since $1,\alpha_0,\ldots,\alpha_0^{p_0-1}$ form a
$\F_0$-basis of $\F$, these traces uniquely determine $c_0$.
Note that, only symbols indexed by $R_0$ contribute to the repair process of $c_0$.
In particular, each symbol provides one element of $\F_0$. 
Note that in contrast to the constructions considered before, the repair parameters in~\cite{TYB17} depend 
on the symbol being repaired. In particular, symbol $i$ is associated with the intermediate field $\F_i$ and 
the repair degree $p_i$.
We also note that, unlike in the constructions considered previously, the evaluation point $0$ is not included in 
the evaluation set.

\subsubsection{Tamo--Barg Codes}\label{preli:LERS:TB}
The repair of Tamo--Barg codes was studied by Sasanka and Lalitha in~\cite{SL22}, 
who proposed three LERS constructions for the more general Tamo--Barg code construction
with local distance $\dTB\geq2$. All schemes use the fact 
that the restriction of the Tamo--Barg code, with locality $r$ and local distance $\dTB,$ 
to each locality set 
is a $[r+\dTB-1,r]_{\F}$ Reed--Solomon code. Consequently, constructing a LERS
for the Tamo--Barg code
reduces to this underlying Reed-Solomon code.

The first construction in~\cite{SL22}, based on~\cite{GW16}, provides a 
local repair scheme for $[n,k,r]_{\F}$ Tamo--Barg codes, with local distance $\dTB,$ whenever 
$q^{m-1}\leq \dTB-1.$

The second construction in~\cite{SL22}, inspired by~\cite{DM17} and \cite{LWJ19}, provides a 
local repair scheme for $[n,k,r]_{\F}$ Tamo--Barg codes,
whenever there exists an integer $a$ such that
$a\mid m,$ $r+\dTB\leq q^a,$ and $q^s\leq \dTB-1.$
Here, the function $g$ and the repair sets are defined using additive cosets, 
analogous to Example~\ref{ex:tb:good_poly_construction} for general local distance $\dTB$.  

Finally, the third construction (inspired by~\cite{TYB17}) shows that
the evaluation points of the locality sets can be chosen such the cut-set bound 
for each underlying Reed--Solomon codes can be applied, 
provided that 
$m \approx(r+\dTB-1)^{(r+\dTB-1)}$.

\subsubsection{Reed--Muller Codes}
The repair scheme in~\cite{GW16} was extended to generalized Reed--Muller codes in~\cite{CZ19}. 
Let $\GRM(\rho,\M)$ be a generalized Reed--Muller code over $\F$. 
Consider a failed node $ \alpha_0=(\alpha_{0,0},\ldots,\alpha_{0,\M-1})\in\F^\M.$ 
The repair scheme is based on the following $m$ check polynomials:
\begin{equation}\label{eq:rm:check-poly} 
	h_i(X)= \prod_{j=0}^{\M-2} \left(1-(X_j-\alpha_{0,j})^{q^m-1}\right) 
	\frac{L_W\bigl(\zeta_i(X_{\M-1}-\alpha_{0,\M-1})\bigr)} {X_{\M-1}-\alpha_{0,\M-1}}, \qquad i\in[m], 
\end{equation}
where $W\subseteq\F$ has dimension $s$ over $\B,$ $L_W$ is a linearized polynomial and $\zeta_1,\ldots
\zeta_m$ is a basis of $\F$ over $\B$.
The first factor ensures that the polynomial vanishes whenever one of the remaining $\M-1$ coordinates differs 
from the corresponding coordinate of the failed node. Thus, the repair problem is effectively reduced to the univariate 
setting in the variable $x_{\M-1}$, where the second factor plays the role of the check polynomials used in the Reed--Solomon construction.
Note that the degree of the check polynomials in~\eqref{eq:rm:check-poly} is at most
    $(\M-1)(q^m-1)+q^s-1\leq \M(q^m-1)-\rho-1 =\rho^\perp.$
Consequently, the check polynomial $h_i$ can be associated to a dual codeword of $\GRM(\rho,\M),$ as required.
Moreover, as shown in~\cite{CZ19}, their satisfy the repair 
conditions~\eqref{eq:gw:basic:m} and~\eqref{eq:gw:basic:1}. 
Hence, they induce a linear exact repair scheme for the failed node $\alpha_0$.
Note that if $\rho=q^m-q^{m-1}-1,$ then $s=m-1.$ In this case, the subspace polynomial $L_W$ reduces to the trace function and the repair 
bandwidth is at most $q^m-1$ over $\B.$
The corresponding leakage functions are then given by 
\begin{equation}\label{eq:rm:leakage_fct}
g_i:c_i\longmapsto
\Tr\left(
	\prod_{j=0}^{\M-2} \left(1-(\alpha_{i,j}-\alpha_{0,j})^{q^m-1}\right) 
	\frac{c_i} {\alpha_{i,\M-1}-\alpha_{0,\M-1}}\right).
\end{equation}
For a single node failure, the construction achieves a repair bandwidth of
$ b=(q^m-1)(m-s)$ over the subfield $\B$, where $ s=\left\lfloor\log_q(q^m-\rho-1)\right\rfloor.$
The above construction requires $q^m-\rho-1>0,$
to ensure that the parameter $s=\left\lfloor\log_q(q^m-\rho-1)\right\rfloor$
is well-defined. Equivalently, the order of the generalized
Reed--Muller code must satisfy 
\begin{equation}\label{eq:rm:cond:lers}
	\rho\leq q^m-2.
\end{equation}
Moreover, since $W$ is an $s$-dimensional $\B$-subspace of
$\F=\F_{q^m}$, it is required that $0\leq s\leq m.$

\renewcommand{\nbase}{n_0}
\renewcommand{\kbase}{k_0}
\renewcommand{\Cbase}{C_0}
\renewcommand{\Gbase}{G_0}
\renewcommand{\Dbase}{D_0}

\section{Deriving LERSs from the Guruswami--Wootters Construction}

In the following, we explore further LERSs that can be obtained from the Guruswami--Wootters construction
in~\cite{GW16} 
for Reed--Solomon codes (see Section~\ref{Ch2::LERS:GW}). 

We first consider Reed--Solomon codes whose dimension is strictly smaller than $q^m-q^{m-1}$. 
In this case, the following proposition shows that the repair of any erased code symbol can be 
performed using fewer than $n-1$ helper nodes, 
resulting in a reduced repair bandwidth.
\begin{proposition}\label{cor:rs:lower_dimensional_result}
Let $\RS(\alpha,k)$ be a Reed--Solomon code of full length (i.e., $n=q^m$) and dimension
$k<q^m-q^{m-1}$
over $\F$. Then there exists a linear exact repair scheme with bandwidth
\[
b=n-1-\left(q^m-q^{m-1}-k\right).
\]
In particular, any $q^m-q^{m-1}-k$ subsymbols are not needed for the repair. 
\end{proposition}

\begin{proof}
Assume that the erased symbol is indexed by
$\alpha_0$. Let $I\subseteq[n-1]$
be an index set of cardinality $|I|=q^m-q^{m-1}-k,$ and define
\[
g(X)=\prod_{i\in I}(X-\alpha_i).
\]
For every message polynomial $f\in\F[X]_{<k}$, it holds that
$\deg(f\cdot g)
\le (k-1)+\deg(g)
< q^m - q^{m-1}.$
Hence, the polynomial $f\cdot g$ satisfies the degree condition required in
Theorem~\ref{thm:gw:LERSexistince}.
Since $g(\alpha_j)=0$ for every $j\in I$, the set of helper nodes can be described by the index set $[n-1]\setminus I,$
whose cardinality is equal to
\[
(n-1)-|I|=
n-1-(q^m-q^{m-1} -k).
\]
Applying Theorem~\ref{thm:gw:LERSexistince} therefore yields a LERS 
with bandwidth $n-1-(q^m-q^{m-1} -k).$
\end{proof}

We next consider a transformation of the check-polynomials associated with an existing LERS.
We show that these polynomials can be transformed by a suitable reversal operation while preserving 
conditions~\eqref{eq:gw:basic:1} and~\eqref{eq:gw:basic:m}. 
\begin{theorem}\label{thm:ex:reverse:LERS}
Consider the $[n,k]_\F$ full-length Reed--Solomon code $\RS(\alpha,k)$, where
$\alpha=(\alpha_0,\ldots,\alpha_{n-1})$ and $q^{m-1}\leq n-k$.
Let $h_1(x),\ldots,h_m(x)$ be the Guruswami--Wootters check-polynomials
$h_i(X)=
\Tr\bigl(\zeta_i(X-\alpha_0)\bigr)/(X-\alpha_0),$ for $i\in[m],$
where $\{\zeta_1,\ldots,\zeta_m\}$ is a $\B$-basis of $\F$.
Define the reversed check-polynomials by
\begin{equation}\label{eq:gw:reverse_check_poly}
    \widetilde{h}_i(X)
=(X-\alpha_0)^{\deg(h_i)}h_i\left(\alpha_0+\frac{1}{X-\alpha_0}\right),\qquad i\in[m].
\end{equation}
Then $\widetilde{h}_1(X),\ldots,\widetilde{h}_m(X)$ form a valid set
of check-polynomials for recovering the symbol that corresponds to the evaluation of $\alpha_0$ with
repair bandwidth at most $n-1$.
\end{theorem}

\begin{proof}
Since
$\deg(h_i)=q^{m-1}-1,$
the assumption $q^{m-1}\leq n-k$ implies
$\deg(\widetilde{h}_i)=q^{m-1}-1<n-k.$
Hence, each $\widetilde{h}_i$ is a valid check-polynomial for the Reed--Solomon code.
Using the definition of $h_i$, we obtain
\begin{align*}
\widetilde{h}_i(x)
&=(x-\alpha_0)^{q^{m-1}-1} h_i\left(\alpha_0+\frac{1}{x-\alpha_0}\right)\\
&=(x-\alpha_0)^{q^{m-1}}\Tr\left(\frac{\zeta_i}{x-\alpha_0}\right)\\
&=\sum_{j=0}^{m-1}\zeta_i^{q^j}(x-\alpha_0)^{q^{m-1}-q^j},
\end{align*}
since $\Tr(z)=\sum_{j=0}^{m-1}z^{q^j}.$
In particular, evaluating at the failed node $\alpha_0$, 
all terms vanish except the term corresponding to $j=m-1$. Thus,
$\widetilde{h}_i(\alpha_0)=\zeta_i^{q^{m-1}}.$
The Frobenius map
$x\mapsto x^{q^{m-1}}$
is a $\B$-linear automorphism of $\F$. Therefore,
$\left\{\widetilde{h}_1(\alpha_0),\ldots,\widetilde{h}_m(\alpha_0)\right\}=
\left\{\zeta_1^{q^{m-1}},\ldots,\zeta_m^{q^{m-1}}\right\}$
is a $\B$-basis of $\F$. Hence,
$\dimB(\spanB(\left\{\widetilde{h}_1(\alpha_0),\ldots,\widetilde{h}_m(\alpha_0)\right\}))=m,$
and condition \eqref{eq:gw:basic:m} is satisfied.
In contrast, for any helper node $\alpha_j\neq\alpha_0$ we have
\begin{align*}
\widetilde{h}_i(\alpha_j)
&=(\alpha_j-\alpha_0)^{q^{m-1}}\Tr\left(\frac{\zeta_i}{\alpha_j-\alpha_0}\right).
\end{align*}
Since 
$\Tr\left(\zeta_i/(\alpha_j-\alpha_0)\right)\in\B,$ it follows that
$\widetilde{h}_i(\alpha_j)
\in
(\alpha_j-\alpha_0)^{q^{m-1}}\B$
for every $i\in[m]$. Therefore,
$\dimB(\spanB(\left\{\widetilde{h}_1(\alpha_j),\ldots,\widetilde{h}_m(\alpha_j)\right\}))\leq 1.$
Thus, at each of the $n-1$ helper nodes condition~\eqref{eq:gw:basic:1} holds.
This proves that the reversed check-polynomials define a valid LERS with bandwidth at most $n-1$.
\end{proof}

Finally, we note that the LERSs in~\cite{GW16} for full-length Reed--Solomon code is invariant under
cyclic permutation of the evaluations points and therefore 
yield further LERSs with the same repair bandwidth.

\begin{proposition}\label{lemma:cyclic_permutation_result}
Consider the full length Reed--Solomon code $RS(\alpha,k)$ over $\F$, where $\alpha=(0,a,a^2,\ldots,a^{n-1})$ for some $a\in\F$ of multiplicative order $n-1$. 
Let $h^{(1)},\ldots,h^{(m)}$ be check-polynomials satisfying
conditions \eqref{eq:gw:basic:1} and \eqref{eq:gw:basic:m}.
For $s\in[n-2]_0$, define
\begin{equation}\label{eq:cyclic_perm}
    \pi_s(0)=0,\qquad\pi_s(i)=i+s\pmod{n-1}\quad\text{for }i\in[n-1].
\end{equation}
Then $\pi_s$ induces an automorphism of $RS(\alpha,k)^\perp$ and the permuted check-polynomials
$\widetilde{h}^{(i)}=\pi_s(h^{(i)})$ for $i\in[m]$ satisfy conditions \eqref{eq:gw:basic:1} and \eqref{eq:gw:basic:m}.
Hence, they define an LERS for the first coordinate with the same repair bandwidth.
\end{proposition}
Proposition~\ref{lemma:cyclic_permutation_result} can be generalized to Reed-Muller codes and follows directly from the 
Theorem 1 in~\cite{KLP68}.

%%%%%%%%%%%%%%%%%%%%%%%%%%%%%%%%%%%%%%%%%%%%%%
%%%%%%%%%%%%%%%%%%%%%%%%%%%%%%%%%%%%%%%%%%%%%%

\section{Systematic Derivation of Leakage via Gr\"obner Bases}\label{Ch5::SystematicDerivationofLERS}
In this chapter, we experimentally analyze their existence for not only Reed--Solomon codes, 
but any linear code (a.k.a. any Massey secret sharing scheme).
We focus in the following on the derivation of LERSs for a given code of length $n$ with bandwidth $n-1$.
The main idea is to reformulate the existence of a LERS as a system of equations over the base field. 
Then, determining all LERSs for a given code is equivalent to solve this reformulated system. 

For this, computational algebra 
techniques, as the Gröbner basis approach, are used. 
For codes with small parameters, Gröbner basis computations are sufficient to determine all solutions. 
However, as the code parameters increase, the resulting systems become computationally challenging to solve. 
To overcome this limitation, we combine the Gröbner basis method with a hybrid approach and a distributed computing framework. 
Together, these techniques yield a fully automated procedure for the systematic derivation of LERSs (resp. leakage functions) 
applicable to any linear code and hence to the corresponding Massey secret-sharing schemes. 
Recall that, (if not other stated) throughout this work we assume that the erased symbol is located at 
the first coordinate of a given codeword.

\subsection{Transformation of the Repair Problem}
Determining all uniform linear exact
repair schemes for a linear $[n,k]_{\F}$ code $C$ with bandwidth
$n-1$ is equivalent to determining all sets of $m$ dual codewords that
satisfy the conditions of Theorem~\ref{thm:gw:LERSexistince}.
Hence, our goal is to determine $m$ dual codewords $h^{(1)},\ldots,h^{(m)}\in C^\perp,$
which we write in a matrix form
    \begin{equation}\label{eq:Hm_matrix}
    H_m = 
    \begin{bmatrix}
    h^{(1)} \\
    h^{(2)}  \\
    \vdots \\
    h^{(m)} 
    \end{bmatrix} =
    \begin{bmatrix}
    h_0^{(1)} & h_1^{(1)} & \cdots & h_{n-1}^{(1)} \\
    h_0^{(2)} & h_1^{(2)} & \cdots & h_{n-1}^{(2)} \\
    \vdots & \vdots &  & \vdots \\
    h_0^{(m)} & h_1^{(m)} & \cdots & h_{n-1}^{(m)}
    \end{bmatrix}.
    \end{equation}
In order to repair the first code symbol, the matrix $H_m$ has to satisfy
\begin{equation}
    \dimB(\spanB(h_0^{(1)},\ldots,h^{(m)}_0) )=m 
\end{equation}
and
\begin{equation}
    \dimB(\spanB(h_j^{(1)},\ldots,h^{(m)}_j) )=1, \qquad \forall j\in[n-1].
    \label{eq:linear_dep_const}
\end{equation}

Express every symbol of the dual
codewords with respect to a fixed $\B$-basis
$\{\zeta_1,\ldots,\zeta_m\}$ of $\F$:
For every coordinate $j$ and every $h^{(\ell)}\in C^\perp$ we write
\[
h_j^{(\ell)}
=
\sum_{i=1}^{m}
h_{j,i}^{(\ell)}
\zeta_i,
\qquad
h_{j,i}^{(\ell)}\in\B.
\]
The bandwidth condition requires that
$\dimB(\spanB(h_j^{(1)},\ldots,h_j^{(m)}))=1,$
for every helper node $j$.
Equivalently, for all $j\in [n-1]$ we have
\[
\rankB\left( 
\begin{bmatrix}
h^{(1)}_{j,1} & \dots & h^{(1)}_{j,m}  \\
\vdots & \ddots & \vdots \\
h^{(m)}_{j,1} & \dots & h^{(m)}_{j,m}
\end{bmatrix}
\right) = 1.
\]
Since every rank-one matrix can be written as an outer product $u_j^T\cdot v_j,$
where $u_j, v_j\in\B^m,$ the matrix $H_m$ can be reformulated as
\[
(H_{m})_\B =
\begin{bmatrix}
&\textbf{1}_{m} & \vline & u_1^T \cdot v_1 & \vline & \cdots & \vline & u_{n-1}^T \cdot v_{n-1}
\end{bmatrix},
\]
where the first block is fixed to the identity matrix to satisfy
the full-rank condition at the erased coordinate. 
This can be done without loss of generality, since 
left multiplication by an invertible $\B$-matrix preserves the 
rank of every block.
Thus, any full-rank matrix at the erased coordinate can be transformed 
into the identity matrix, yielding an equivalent LERS. 
Allowing other full-rank matrices would therefore only introduce equivalent 
representations and would not increase the number of distinct LERSs.
It remains to ensure that the rows of $H_m$ are valid dual codewords. 
This is equivalent to requiring
$G \cdot H_m^T = \textbf{0}_{k\times m},$
where $G$ denotes a generator matrix of $C$.
Expanding this matrix equation with respect to the chosen basis of $\F$
yields a polynomial system over the base field $\B$: 
\begin{equation}
    (G \cdot H_m^T)_\B = \textbf{0}_{mk\times m},
    \label{eq:linear_system_gene}
\end{equation}
where the right expression in \eqref{eq:linear_system_gene} is the matrix obtained 
by expanding each entry of $G \cdot H_m^T$ with respect
to the $\B$-basis $\zeta_1,\ldots\zeta_m,$ column-wise. 
Solving \eqref{eq:linear_system_gene}
is therefore equivalent to determining all LERSs with bandwidth at most $n-1$, i.e., in which each helper node sends one $\B$-symbol.

\subsection{Solving the System of Equations}
The polynomial system, described in~\eqref{eq:linear_system_gene}, is solved using Gröbner basis
techniques. A Gröbner basis, first introduced in~\cite{B65}, 
defines (for a given monomial ordering) a canonical generating set of a polynomial ideal,
that allows the solution set of multivariate polynomial systems to be
determined algorithmically. The Buchberger algorithm computes such a 
basis by successively transforming and reducing the input polynomials, 
replacing them with an equivalent generating set with suitable structural properties. 
As demonstrated in the \textsc{Magma} Handbook (Example \texttt{GB\_SolveOverGF2(H115E5)}), 
using the Gr\"obner basis approach,
to solve the system of equations given in~\eqref{eq:linear_system_gene} 
over $\B$ does not automatically restrict the solutions to $\B$. 
Hence, solutions in the algebraic closure of the 
extension field $\F$ are also obtained, which strongly increases the size of the solution 
space and results in a larger running time. 
Since we are only interested in solutions whose entries lie in the base 
field $\B$, we restrict the solution set by adding the corresponding 
\emph{field equations} to the system. 
More formally, let
$\B[U_{j,i},V_{j,i} : (j,i)\in[n-1]\times[m]]$
denote the polynomial ring over $\B$, where $U_{j,i}$ and $V_{j,i},$ 
for $(j,i)\in[n-1]\times[m],$ are indeterminates. We solve the following 
algebraic system:
\begin{equation}\label{eq:grobner_basis_algebraic_system}
    \left\{
    \begin{aligned}
    (G \cdot H_m^T)_\B &= \mathbf{0}_{mk\times m},\\
    U_{j,i}^q-U_{j,i} &= 0
    && \text{for all } (j,i)\in[n-1]\times[m],\\
    V_{j,i}^q-V_{j,i} &= 0
    && \text{for all } (j,i)\in[n-1]\times[m],
    \end{aligned}
    \right.
\end{equation}
which is defined over the polynomial ring $\B[U_{j,i},V_{j,i}].$
Then a solution to this system assigns concrete values $u_{j,i},v_{j,i}\in\B$ 
to the indeterminates $U_{j,i}$
and $V_{j,i}$, respectively, such that all equations in the system are satisfied.
Note that for~\eqref{eq:grobner_basis_algebraic_system}, the Gröbner basis technique 
yields \textit{all solutions} in $\B$.

\begin{table}[h]
    \centering
    \captionsetup{width=\textwidth}
    \setlength{\tabcolsep}{8pt}
    \begin{tabular}{
                    c  % Time (string like 01:23:45)
            c  % q
            c  % m
            c  % k
            c  % Solutions
    }
            \toprule
            {Time (sec)} & {$q$} & {$m$} & {$k$} & {\#LERSs} \\
            \midrule
            0.130  & 2 & 2 & 1  & 60 \\    
            2.160  &2 & 2 & 2  & 3 \\ 
            \bottomrule
    \end{tabular}
    \caption{Experimental results for full-length Reed--Solomon codes over the field
    of size $q^m$ using the Gröbner basis technique on a machine with 20 CPU and 32 GiB.}
    \label{tab:summary_non-hybrid_approach}
\end{table}

First results were successfully obtained by solving the system of equations 
in~\eqref{eq:grobner_basis_algebraic_system} using the Gröbner
basis technique (see Table~\ref{tab:summary_non-hybrid_approach}).
Solving~\eqref{eq:grobner_basis_algebraic_system}
 yields all solutions over the base field, and
these solutions are aligning with those obtained by Guruswami and
Wootters. 
% Table~\ref{tab:summary_non-hybrid_approach} summarizes the first
% experimental results obtained by successfully solving the system of
% equations in~\eqref{eq:grobner_basis_algebraic_system} using the Gröbner
% basis technique. 
The experiments were performed on a machine with 
$20$ CPU cores and $32$ GiB of available RAM.
However, with the current approach and the available 
computational resources,
we were unable to solve instances
with larger parameters than those reported in
Table~\ref{tab:summary_non-hybrid_approach}. 
\begin{table}[h]
    \centering
    \captionsetup{width=\textwidth}
    \setlength{\tabcolsep}{5pt}
    \begin{tabular}{
                    c  % Time (string like 01:23:45)
            c  % q
            c  % m
            c  % k
            c  % Solutions
    }
            \toprule
            {Time (sec)} & {$q$} & {$m$} & {$k$} & {\#LERSs} \\
            \midrule
            $198(\approx3.3$ \text{min.}) & 2 & 3 & 4  & 14 \\
            \greyMidrule
            $1020141(\approx 11.8$ \text{days}) &  3 & 2 & 5  & 292 \\
            47.82  & 3 & 2 & 6  & 4 \\
            \bottomrule
    \end{tabular}
    \caption{Experimental results for full-length Reed--Solomon codes over the field
    of size $q^m$ using the Gröbner basis technique on a machine with 512 CPU 
    and 1024 GiB.}
    \label{tab:summary_non-hybrid_approach_magarat}
\end{table}

In a second set of experiments, we used the Margaret high-performance 
computing
cluster, which provides $512$ CPU cores and $1024$ GiB of RAM. 
This increased the
range of parameters for which the system could be successfully solved, 
as shown
in Table~\ref{tab:summary_non-hybrid_approach_magarat}. 
However, the computational
limits were again reached quickly due to the rapidly grow of the computational complexity of the Gröbner basis method
with the number of
variables. 
Consequently, solving the systems that arise from larger codes
becomes computationally infeasible.
For this reason, both the hybrid approach 
and a distributed computing framework were integrated into the program in order to improve the efficiency.

\subsection{Hybrid Approach}\label{sec:hybrid_approach}

To reduce the computational complexity, a hybrid strategy is employed~\cite{BFP09}.
The original system contains $2m(n-1)$
$\B$-unknowns, corresponding to the vectors
$ U_j,V_j$ for every helper node.
The main idea of the hybrid approach is to fix a subset of these unknowns
in advance to then solve the resulting smaller systems independently.
This reduces the number of indeterminates in each smaller system, thereby making 
the Gröbner basis computations computationally more feasible.
However, this comes at the cost of having to compute Gröbner bases 
for multiple smaller systems. Thus, the hybrid approach trades the 
computational cost of a single large Gröbner basis computation for 
multiple smaller computations, balancing the size of the individual 
systems against the number of systems that need to be solved.
\begin{algorithm}[h]
\caption{Hybrid Approach}
\label{alg:hybrid_approach}
\begin{algorithmic}[1]
\Require Algebraic system~\eqref{eq:grobner_basis_algebraic_system}, 
$\B,\F$, set of fixed unknowns
$S\subseteq \{U_{j,i},V_{j,i}\}$
\Ensure All solutions of~\eqref{eq:grobner_basis_algebraic_system}
\State Enumerate all possible $q^{|S|}$ assignments of $\B$-values to the unknowns in $S$.
\State $\mathtt{sol} \leftarrow \emptyset$
\For{each assignment}
\State Substitute the assigned values into~\eqref{eq:grobner_basis_algebraic_system}
\State $\mathtt{current\_sol} \leftarrow$ solve the resulting system using Gröbner approach
\State $\mathtt{sol} \leftarrow \mathtt{sol} \cup \mathtt{current\_sol}$
\EndFor
\State \Return $\mathtt{sol}$
\end{algorithmic}
\end{algorithm}
Algorithm~\ref{alg:hybrid_approach} illustrates the structure of the hybrid approach for a given 
subset $S$ of fixed unknowns. 
We are focusing on two different settings. We either fix $n-1$ variables or $2(n-1).$
For every column $j$ in \eqref{eq:Hm_matrix}, we fix the first coordinate
$U_{j,1}$ (resp. two coordinates $U_{j,1}$ and $U_{j,2}$) of the vector $U_j=(U_{j,1},\ldots,U_{j,m})$, such that 
the system~\eqref{eq:grobner_basis_algebraic_system} needs to be solved over 
$$\B[U_{j,i},V_{j,i} : (j,i)\in[n-1]\times[m], i\neq1; V_{j,1}: j\in [n-1]],$$ 
resp. 
$$\B[U_{j,i},V_{j,i} : (j,i)\in[n-1]\times[m]i\neq1,2; V_{j,1}, V_{j,2} : j\in [n-1]]$$ 
instead of $\B[U_{j,i},V_{j,i} : (j,i)\in[n-1]\times[m]].$
This reduces the number of unknowns to $(2m-1)(n-1)$ (resp. $(2m-2)(n-1)$).
Since each unknown can take any value in $\B$, there are initially
$q^{n-1}$ (resp. $q^{2(n-1)}$) possible assignments.

This space can be reduced further by exploiting the minimum distance 
of the dual code. Recall that every nonzero dual codeword $h=(h_0,\ldots,h_{n-1})\in C^\perp$
has Hamming weight at least $d_{H}^\perp$, where $d_H$ denotes the minimum Hamming distance. 
Equivalently, every nonzero 
dual codeword contains at most $n-d_{H}^\perp$
zero coordinates.
Since fixing $u_{j,1}=0$ (resp. $u_{j,2}=0$) forces the $j$-th coordinate of $h^{(1)}$ (resp. $h^{(2)}$) 
to be zero, at most
$n-d_{H}^\perp$ of the values $u_{1,1},\ldots,u_{n-1,1}$ (resp. $u_{1,2},\ldots,u_{n-1,2}$)
can be assigned to zero. Consequently, it is sufficient to consider 
only those assignments in which at most 
$n-d_{H}^\perp$ values are fixed to zero.
Hence, the number of polynomial systems that must be solved using $n-1$ fixed unknowns is reduced from
$q^{n-1}$ to
\begin{equation}\label{eq:number_teask_n-1}
    \sum_{i=0}^{n-d_{H}^\perp}
    \binom{n-1}{i}(q-1)^{n-1-i}.
\end{equation}
Equivalent, for $2(n-1)$ fixed unknowns the number of polynomial systems is reduced to 
\begin{equation}\label{eq:number_task_2(n-1)}
    \left(
    \sum_{i=0}^{n-d_{\min}^\perp}
    \binom{n-1}{i}(q-1)^{n-1-i}
    \right)^2.
\end{equation}
\begin{example}
    Consider a MDS $[16,8]_\F$ code, where the finite field $\F$ has size $2^4.$ 
    Then using $n-1$ fixed unknowns the number of polynomial systems is reduced 
    from $32,768$ to 
    $16,384.$
    Equivalently, using $2(n-1)$ fixed unknowns the number of polynomial systems is reduced from 
    $1,073,741,824$ to
    $268,435,456.$
\end{example}

\subsection{Experimental Setting}
To handle the large number of independent systems arising in the hybrid approach, we employed the manager–worker 
architecture provided by \textsc{Magma}'s distributed system~\cite{Magma}. In this setup, a single manager process controls the workflow: 
it generates the systems of equations, assigns them to worker processes, and collects the corresponding results. 
The worker processes are responsible for solving the assigned systems and returning the solutions to the manager.
This model is particularly well suited for our application, since the systems are independent and can be solved in 
parallel without internal worker communication. The manager–worker scheme in \textsc{Magma} ensures efficient load balancing 
by dynamically distributing tasks to available workers. Consequently, the approach achieves a substantial reduction 
in wall-clock time while keeping the implementation conceptually simple and robust.

For instance, for the parameters $q=2$, $t=4$, $k=8$ and $n=16$, the program 
requires a wall-clock time of $426$ seconds, using  with $256$ workers (on a computer with 256 cores), 
while the total time is 39,300 seconds. 
This corresponds to a reduction of approximately $98.9 \%$ of wall-clock time.

\subsection{Program Summary and First Examples}
\begin{figure}[h]
    \centering
    \includegraphics[width=0.9\linewidth]{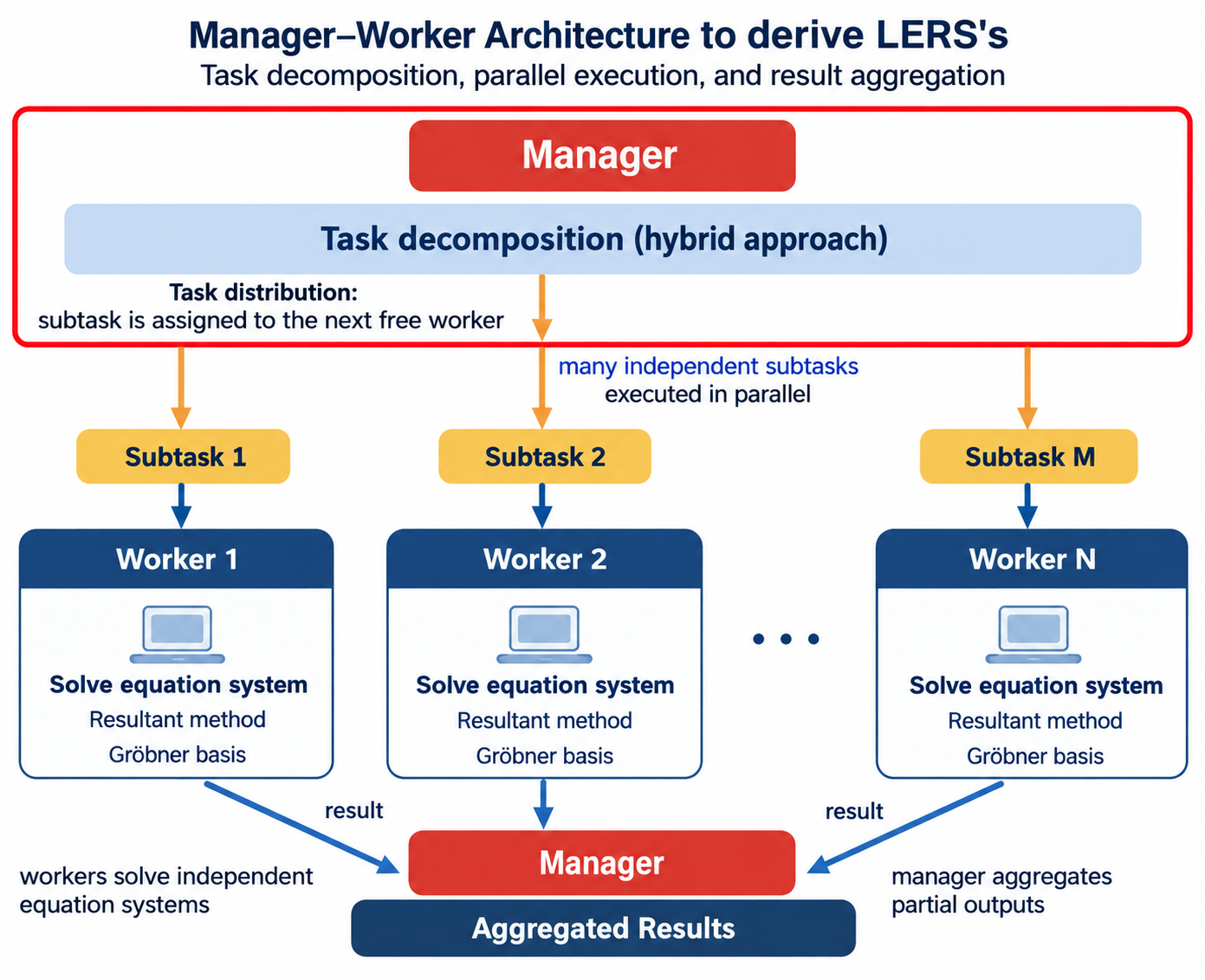}
    \caption{Overview about the program structure.}
    \label{fig:Overview_program}
\end{figure}
Figure~\ref{fig:Overview_program} provides an overview of the resulting program structure, 
incorporating the building blocks described before. 
The number of subtasks is determined by either the number of fixed values and results for 
$n-1$ (resp. $2(n-1)$) fixed values into \eqref{eq:number_teask_n-1} (resp. \eqref{eq:number_task_2(n-1)}) subtask. 
To further improve the efficiency of the program, the resultant-based method \cite{CLS05} can optionally be integrated 
into the computation. 
It eliminates selected indeterminates by combining polynomial equations 
through resultants, thereby reducing the number of indeterminates in the 
system before applying the Gröbner basis computation.
However, we observed that the use of the resultant method does not necessarily 
improve the overall performance in all cases and is therefore not always be considered.
Note that the program is applicable to any linear code by appropriately adapting the generator matrix. 
In the following, we will present some direct applications of the program.
\begin{example}[Binary Reed--Solomon code]
Consider the finite field $\F=\mathbf{F}_{2^2}$ and the Reed--Solomon code $\RS(\alpha,2)_\F$ of length $4$, with evaluation points
$\alpha=(0,1,a,a+1),$ where $a$ is a primitive element of $\F$.
Applying our program to this code yields the following three distinct sets of dual codewords:
$
\{(1,a+1,a,0),(a,a+1,0,1)\},
\{(1,0,a+1,a),(a,1,a+1,0)\},
\{(1,a,0,a+1),(a,0,1,a+1)\}.
$
% $$
% \begin{aligned}
% &\{(1,a+1,a,0),(a,a+1,0,1)\},\\
% &\{(1,0,a+1,a),(a,1,a+1,0)\},\\
% &\{(1,a,0,a+1),(a,0,1,a+1)\}.\\
% \end{aligned}
% $$
% These sets correspond, respectively, to the following sets of polynomials:
% $$
% \begin{aligned}
% &\{a x+1,\;x+a\},\\
% &\{x+1,\;a^2x+a\},\\
% &\{a^2x+1,\;ax+a\}.
% \end{aligned}
% $$
% For comparison, applying the construction of \cite{GW16} to the same code $\RS(\alpha,2)_\F$ yields the set
% ${(1,0,a+1,a),(a,1,a+1,0)},$ which coincides with one of the three solutions identified by our program. 
For comparison, we next apply the construction of \cite{GW16} to the same code $\RS(\alpha,2)_\F$. 
Let $\{1,a\}$ be a $\B$-basis of $\F$, and let $\alpha_0=0$ denote the evaluation point in $\alpha$ corresponding to the 
coordinate to be repaired. The GW construction considers the check polynomials
$$
p_1(X)=\frac{\Tr(X)}{X}=1+X,
\qquad
p_2(X)=\frac{\Tr(aX)}{X}=a+a^2X.
$$
Evaluating these polynomials at the evaluation points and applying the corresponding dual multipliers yields the dual codewords
$$
\begin{aligned}
 h_1=(1,0,a+1,a),
\qquad
 h_2=(a,1,a+1,0).
\end{aligned}
$$
Hence, the GW construction produces the set
$\left\{(1,0,a+1,a),\,(a,1,a+1,0)\right\},$
which coincides with one of the three solutions identified by our program.
Note that the other two solutions can be obtained from the GW construction by 
cyclic permutation, confirming Proposition~\ref{lemma:cyclic_permutation_result}. 
Our program does solve the complete system in $0.04$ seconds of wall-clock time, provided that we use $20$ workers and 
fix $n-1$ variables.
\end{example}
\begin{example}[Non-binary Reed--Solomon code]
    Consider the finite field $\F=\mathbf{F}_{3^2}$ and the code $\RS(\alpha,6)_\F$ of length $9$, 
    with evaluation points
    $\alpha=(0,1,a,\ldots,a^7),$
    where $a$ is a primitive element of $\F$ satisfying $a^2+a+2=0.$
    Our program finds four distinct sets of dual codewords, one of which coincides with the repair 
    scheme obtained from the construction of \cite{GW16}. The remaining solutions are obtained by 
    cyclic permutation, in accordance with Proposition~\ref{lemma:cyclic_permutation_result}. Only four distinct sets arise, 
    since cyclic shifts by $i$ and $i+4$ yield identical sets for all $i\in[4]$.
\end{example}

\section{Experimental Results}

This section presents the experimental results obtained by applying the program 
introduced above to various linear codes with different parameter settings. 
The experiments were performed on the Margaret high-performance computing cluster, 
which provides 512 CPU cores and $1024$ GiB of RAM.
Note that, in all experiments, the number of CPU cores is chosen to match the number of workers, 
such that each worker is assigned a dedicated CPU core. 
Recall, the program solves the algebraic system~\eqref{eq:grobner_basis_algebraic_system} 
and therefore computes all sets of dual codewords that 
induce a uniform LERS with bandwidth at most $n-1$.
We also considered an additional simplification, controlled by the 
parameter $\ell$. More precisely, we considered the case in which the last $\ell$ columns (a.k.a. positions) are fixed to zero, 
which further reduces the number of unknowns and allows us to search 
for LERSs with bandwidth at most $n-1-\ell$.
The column $T$ report 
the wall-clock running time for the completion of the program.
All running times are given in seconds. 
A dash indicates that the corresponding experiment was either not performed or did not terminate successfully.
The column ``LERS“ gives the number of LERSs identified by the program. Each distinct set of valid dual 
codewords is counted as one LERS. Finally, the ``Status“ column, if given, indicates whether the 
computation terminated successfully (reported with ``Completed“) or was terminated prematurely, 
for example due to resource limitations (reported with ``Terminated“). If the ``Status“ 
column is omitted then it can be assumed that all 
displayed experiments successfully completed.

%%%%%%%%%%%%%%%%%%%%%%%%%%%%%%%%%%%%%%%%%%%%
%%%%%%%%%%%%%%%%%%%%%%%%%%%%%%%%%%%%%%%%%%%%

\subsection{Reed--Solomon Codes}\label{subsec:RS_Grobner}

We first consider Reed--Solomon codes. Let $\F$ be a finite field of size $q^m$, and consider a Reed--Solomon code $\RS(\alpha,k)_\F$ of 
length $n.$ Throughout our experiments for Reed--Solomon codes, 
we use the evaluation points
$\alpha=(0,1,a,a^2,\ldots,a^{n-2}),$
where $a\in\F$ is a primitive element of $\F$. We assume that the first code coordinate, 
corresponding to the evaluation point $\alpha_0=0$, is erased and has to be repaired.

In the following, selected experimental results are summarized. 
A complete summary of the experimental result can be found in the Appendix~\ref{Appendix:exper_results}. 

Firstly, we noticed a direct relationship 
between the existence of 
LERSs and the dimension of Reed--Solomon codes of length $n=m+1$. 
In particular, the observed behavior indicates that the existence of 
low-bandwidth LERSs may be determined by the code dimension. 
The corresponding experimental results are summarized in 
Table~\ref{table:RS:k:1}.

\begin{table}[h]
    \centering
    \setlength{\tabcolsep}{4pt}

    % Left table
    \begin{minipage}[t]{0.45\textwidth}
        \centering
        \begin{tabular}{cccccc}
            \toprule
            $q$ & $m$ & $n$ & $k$ & $\ell$ & LERSs \\
            \midrule
            2 & 2 & 3 & 1 & 0 & 6 \\
            \midrule
            2 & 3 & 4 & 1 & 0 & 168 \\
            2 & 3 & 4 & 2 & 0 & 0 \\
            \midrule
            2 & 4 & 5 & 1 & 0 & 20160 \\
            2 & 4 & 5 & 2 & 0 & 0 \\
            2 & 4 & 5 & 3 & 0 & 0 \\
            \bottomrule
        \end{tabular}
    \end{minipage}
    \hspace{2.5cm}
    % Right table
    \begin{minipage}[t]{0.45\textwidth}
        \centering
        \begin{tabular}{cccccc}
            \toprule
            $q$ & $m$ & $n$ & $k$ & $\ell$ & LERSs \\
            \midrule
            2 & 5 & 6 & 1 & 1 & 0 \\
            2 & 5 & 6 & 2 & 0 & 0 \\
            2 & 5 & 6 & 3 & 0 & 0 \\
            \midrule
            3 & 3 & 4 & 1 & 0 & 1404 \\
            3 & 3 & 4 & 2 & 0 & 0 \\
            \bottomrule
        \end{tabular}
    \end{minipage}

    \caption{Experimental results for Reed--Solomon codes
    of length $n=m+1$ over the field of size $q^m$
    with $n-1$ fixed variables.}
    \label{table:RS:k:1}
\end{table}
We observe and proof, that for Reed--Solomon codes of length $n=m+1$, a LERS with 
bandwidth at most $n-1$ exists if and only if $k=1$.
\begin{proposition}\label{cor:RSone-dimLERS}
Let $\RS(\alpha,k)\subseteq\F^n$ be a Reed--Solomon code with $n=m+1$. Then $\RS(\alpha,k)$ admits a linear exact repair scheme 
with bandwidth at most $n-1$ if and only if $k=1$.
\end{proposition}
\begin{proof}
For $k=1$, the code is one-dimensional, and any erased code symbol can be recovered 
by only querying one sub symbol per surviving node. 
Hence, a LERS with bandwidth at most $n-1$ exists.

For the converse, suppose that $k>1$. Since $n=m+1$, we have $n-1=m$ available helper nodes. 
By the cut-set bound for exact repair in~\eqref{eq:bandwidth:cut-set-bound}, the repair bandwidth $b$ 
of any MDS code over an extension field with 
extension degree $m$ satisfies 
$$
b \geq \frac{m^2}{m-k+1}>m=n-1.
$$
The second inequality follows from $m-k+1<m$, which holds since $k>1$.
Thus, no LERS with bandwidth at most $n-1$ can exist for $k>1$, which proofs the claim.
% Therefore, a LERS with bandwidth at most $n-1$ exists if and only if $k=1$.
\end{proof}
Note that, Proposition \ref{cor:RSone-dimLERS} holds for any MDS code.

\begin{table}[h]
    \centering
    \setlength{\tabcolsep}{4pt}

    % Left table
    \begin{minipage}[t]{0.45\textwidth}
        \centering
        \begin{tabular}{ccccc}
            \toprule
            $q$ & $m$ & $n$ & $k$ & LERSs \\
            \midrule
            2 & 2 & 3 & 1 & 6 \\
            2 & 2 & 4 & 2 & 3 \\
            \midrule
            2 & 3 & 7 & 3 & 28 \\
            2 & 3 & 8 & 4 & 14 \\
            \bottomrule
        \end{tabular}
    \end{minipage}
    \hspace{2.5cm}
    % Right table
    \begin{minipage}[t]{0.45\textwidth}
        \centering
        \begin{tabular}{ccccc}
            \toprule
            $q$ & $m$ & $n$ & $k$ & LERSs \\
            \midrule
            2 & 4 & 15 & 7 & 60 \\
            2 & 4 & 16 & 8 & 30 \\
            \midrule
            3 & 2 & 8 & 5 & 8 \\
            3 & 2 & 9 & 6 & 4 \\
            \bottomrule
        \end{tabular}
    \end{minipage}

    \caption{Experimental results for Reed--Solomon codes
    of length $n$ over the field of size $q^m$
    with $\ell=0$.}
    \label{table:RS:double:LERS}
\end{table}
Next, the experiments reported in Table~\ref{table:RS:double:LERS} suggest that, 
for full-length Reed--Solomon codes with $k=q^m-q^{m-1}$, the number of 
LERSs is doubled when both the length and the dimension of the code are 
reduced by one. More formally:
\begin{observation}
    The experimental results reported in Table~\ref{table:RS:double:LERS} strongly suggest that the number 
    of LERSs for a full-support Reed--Solomon code $\RS(\alpha,q^m-q^{m-1})_\F$ is twice the number of LERSs 
    for the shortening of $\RS(\alpha,q^m-q^{m-1})_\F$ in the last position (corresponding to the 
    evaluation point $a^{q^m-2}$), 
    namely $\RS(\alpha,q^m-q^{m-1}-1)_\F$ of length $n=q^m-1$.
\end{observation}

We further observe that forcing $\ell$ columns to be zero is analogous to 
shortening a Reed--Solomon code. The corresponding results are summarized 
in Table~\ref{table:RS:zero-columns} and are consistent with 
Proposition~\ref{cor:rs:lower_dimensional_result}.
Furthermore, these results suggest the following proposition which holds for any linear code:
\begin{proposition}
Let $C$ be a $[n,k]_\F$ code, and let $I\subseteq [n]$ be an index set with $|I|=\ell$. 
Then, LERSs whose corresponding dual codewords have zero columns at the positions in $I$ are 
in one-to-one correspondence with LERSs for the Reed--Solomon $[n-\ell,k]$ code $\overline{C}_I$ obtained by shortening $C$ at 
the positions in $I$.
\end{proposition}

\begin{proof}
Since the columns indexed by $I$ are zero, these positions do not contribute to the corresponding dual codewords. 
Removing these columns therefore gives the same repair conditions on the remaining $n-\ell$ positions. 
The resulting code is the shortening of $C$ at the positions in $I$, and hence the corresponding 
LERSs are in one-to-one correspondence.
\end{proof}

\begin{table}[h]
    \centering
    \setlength{\tabcolsep}{4pt}

    % Left table
    \begin{minipage}[t]{0.38\textwidth}
        \centering
        \begin{tabular}{cccccc}
            \toprule
            $q$ & $m$ & $n$ & $k$ & $\ell$ & LERSs \\
            \midrule
            2 & 3 & 6 & 2 & 0 & 686 \\
            2 & 3 & 8 & 2 & 2 & 686 \\
            \midrule
            3 & 2 & 6 & 1 & 1 & 14580 \\
            3 & 2 & 6 & 2 & 0 & 5868 \\
            3 & 2 & 7 & 4 & 0 & 20 \\
            3 & 2 & 8 & 1 & 3 & 14580 \\
            3 & 2 & 8 & 2 & 2 & 5868 \\
            \bottomrule
        \end{tabular}
    \end{minipage}%
    \hspace{2.4cm}%
    % Right table
    \begin{minipage}[t]{0.38\textwidth}
        \centering
        \begin{tabular}{cccccc}
            \toprule
            $q$ & $m$ & $n$ & $k$ & $\ell$ & LERSs \\
            \midrule
            3 & 2 & 8 & 3 & 1 & 2328 \\
            3 & 2 & 8 & 4 & 1 & 20 \\
            3 & 2 & 8 & 4 & 0 & 896 \\
            3 & 2 & 9 & 1 & 4 & 14580 \\
            3 & 2 & 9 & 2 & 3 & 5868 \\
            3 & 2 & 9 & 3 & 2 & 2328 \\
            3 & 2 & 9 & 4 & 1 & 896 \\
            \bottomrule
        \end{tabular}
    \end{minipage}

    \caption{Experimental results for Reed--Solomon codes
    of length $n$ over the field of size $q^m$,
    with $n-1$ fixed variables.}
    \label{table:RS:zero-columns}
\end{table}

\subsection{Generalized Reed--Muller Codes}
Let us now consider Reed--Muller codes. For this, consider a generalized Reed--Muller 
code $\GRM(\rho,\M)$ over $\F$ of length $n=(q^m)^\M$, with generator matrix $G$. 
Again, the ordering of the columns of $G$ determines the ordering of the code coordinates and, equivalently, of the evaluation points. 
Throughout our experiments, we evaluate the code at all points of $\F^\M$.
More precisely, let $a\in\F$ be a primitive element and fix the ordering
$
\beta=(0,1,a,a^2,\ldots,a^{q^m-2})
$
of the elements of $\F$. We order the evaluation points according to the Cartesian product
$$
\F^\M
=
\underbrace{\beta\times\cdots\times\beta}_{\M\text{ times}},
$$
where we start by varying the first component over all elements of $\F$, then the second component, and so on. 
Thus, for $\M=2$, the ordering begins with
$
(0,0),(1,0),(a,0),\ldots,(a^{q^m-2},0),$
$(0,1),(1,1),(a,1),\ldots,(a^{q^m-2},1),
\ldots,
$
and ends with $(a^{q^m-2},a^{q^m-2}).$
In particular, the first evaluation point in the general setting is
$
\alpha_0=(0,\ldots,0).
$
In the following, the first code symbol, corresponding to the evaluation point $\alpha_0$, is erased 
and has to be repaired.

\begin{table}[h]
    \centering
    \setlength{\tabcolsep}{8pt}
    \begin{tabular}{ccccccccc}
        \toprule
        $q$ & $m$ & $n$ & $\rho$ & $\M$ &
        Workers &
        LERSs &
        Status \\
        \midrule

        2 & 2 & 4 & 0 & 1 & 20  & 60 & Completed \\
        2 & 2 & 4 & 1 & 1 & 20   & 3 & Completed \\
        2 & 2 & 16 & 1 & 2 & 128  & $\geq0$ & Terminated \\
        2 & 2 & 16 & 1 & 3 & 256  &  0& Completed \\
        2 & 2 & 16 & 2 & 2 & 256   &  0& Completed \\
        2 & 2 & 16 & 3 & 2 & 128  & $\geq 491$ & Terminated\\
        2 & 2 & 16 & 3 & 3 & 128  &0  & Completed \\
        2 & 2 & 64 & 3 & 3 & 128  &0  & Completed \\

        \midrule

        3 & 2 & 16 & 6 & 2 & 128 &  0& Completed \\
      
        \bottomrule
    \end{tabular}
    \caption{Computational results for generalized Reed--Muller codes over the field
    $\mathbb{F}$ with $n-1$ fixed variables.}
    \label{table:RM}
\end{table}
Table~\ref{table:RM} summarizes selected computational results obtained
by applying the program 
to the just described set-up for Reed--Muller codes. Note that, the derived LERS for a generalized Reed--Muller code
$\GRM(\rho,\M)$ over $\F$ have bandwidth $b\leq(q^m)^\M-1.$
For $\M=1$, the Reed--Muller codes reduce to Reed--Solomon codes, and
the obtained results are therefore consistent with the corresponding
Reed--Solomon results.
Consider $\GRM(3,2)$ over
$\F=\mathbf{F}_{2^2}$, which has length $ n=(q^m)^\M=4^2=16.$
Note that the parameter condition~\eqref{eq:rm:cond:lers} required by
the repair scheme construction in~\cite{CZ19} is not satisfied. Consequently, the
construction in~\cite{CZ19} does not provide a LERS for this parameter setting.
Nevertheless, our program systematically derivates $491$ LERSs different
for the code $\GRM(3,2)$.
Among these LERSs, we find schemes that do not require information from
all helper nodes. In particular, three helper nodes can be omitted
entirely while still allowing the erased symbol to be recovered. This
results in an LERS with repair bandwidth $ b=12.$

\subsection{Tamo--Barg Codes}
While applying the program to Tamo--Barg codes $\TB(n,k,r)$ over a finite field $\F$, 
we order the evaluation points according to their local repair groups.
More precisely, let $r+1$ denote the size of each local group, and let $a\in\F$ be a primitive element. 
The evaluation points are grouped into consecutive blocks of $r+1$ points. 
For $i\in[n/(r+1)-1]_0$, the $i$-th local repair group consists of
$$
\left(
a^i,\,
a^{i+(q^m-1)/(r+1)},\,
a^{i+2(q^m-1)/(r+1)},\,
\ldots,\,
a^{i+r(q^m-1)/(r+1)}
\right).
$$
Thus, each consecutive block of $r+1$ evaluation points corresponds to one local repair group.
The first symbol, 
which is assumed to be erased, corresponds to the evaluation point $\alpha_0=1$.
\begin{table}[h]
    \centering
    \setlength{\tabcolsep}{8pt}
    \begin{tabular}{ccccccccc}
        \toprule
        $q$ & $m$ & $n$ & $k$ & $r$ & $\ell$ &
       $T$ & LERSs & Status \\
        \midrule

        2 & 4 & 15 & 5 & 4 & 3  & -- & $\geq 95$ & Terminated \\
       % 2 & 4 & 15 & 5 & 4 & 4 &  -- & -- & 0 & Completed \\
        2 & 4 & 15 & 6 & 4 & 0  & -- & $\geq$ 20 & Terminated \\
        2 & 4 & 15 & 6 & 4 & 3  & 802.72 & 18 & Completed \\
        2 & 4 & 15 & 7 & 4 & 0  & -- & $\geq$ 0 & Terminated \\
        2 & 4 & 15 & 7 & 4 & 3 & 8 & 0 & Completed \\
        2 & 4 & 15 & 8 & 4 & 0  & 158 & 0 & Completed \\
        \midrule
        3 & 2 & 8 & 4 & 3 & 0 & 5.15 & 700 & Completed\\
        3 & 2 & 8 & 4 & 3 & 2 & 0.95 & 9 & Completed\\
        3 & 2 & 8 & 5 & 3 & 2 & 1.67 & 0 & Completed \\
        \bottomrule
    \end{tabular}
    \caption{Experimental results for Tamo--Barg codes over the field
    $\mathbb{F}$ with $n-1$ fixed variables and 256 workers.}
    \label{table:TB}
\end{table}
Our experiments for the parameters shown in Table~\ref{table:TB} reveal that 
the Tamo--Barg codes $\TB(15,6,4)$, $\TB(15,5,4)$ over $\mathbf{F}_{2^4}$
and $\TB(8,4,3)$ over $\mathbf{F}_{3^2}$ admit LERSs with bandwidth 
$n-1-\ell= 15-1-3=11$ and $n-1-\ell= 8-1-2=5$.
Note that in all cases, the corresponding LERSs are not covered by the 
construction in~\cite{SL22} (see section~\ref{preli:LERS:TB}) and therefore describe new constructions of
LERS.

\subsection{Further Algebraic Properties} 
Observed patterns in experimental results motivated us to formulate and proof further algebraic properties
that apply to arbitrary linear codes. 
\begin{proposition}\label{cor:one-dimLERS}
Let $C$ be a $[n,k]_{\F}$ code with $n=m+1$.
Then there exists a linear exact repair scheme with bandwidth $n-1$
if $k=1$.
\end{proposition}

\begin{proof}
Assume that a LERS with bandwidth $b=n-1=m$ exists.
By the Guruswami--Wootters characterization, there exist
$m$ dual codewords $h^{(1)},\ldots,h^{(t)}\in C^\perp$
such that condition \eqref{eq:gw:basic:1} and \eqref{eq:gw:basic:m} are fulfilled.
Since the $m$ dual codewords are linearly independent over $\F$ (see \eqref{eq:gw:basic:m}), we have
$n-k=\dimB(C^\perp)\ge m.$
Using $n=m+1$, we obtain $m+1-k\ge m,$
which is equivalent to $k\le1.$
Since every non-trivial linear code (i.e., a code different from $\{0\}$) satisfies $k\ge1$, it follows that
$k=1$.
\end{proof}
Note that Proposition~\ref{cor:one-dimLERS} becomes an if-and-only-if statement
for MDS codes 
(see Proposition~\ref{cor:RSone-dimLERS}). 

Next, we show that the existence of a LERS does
induce a constraint on the dimension $k$ of MDS codes of length strictly smaller than
$q^m.$ 
\begin{theorem}
Let $C\subseteq\F^n$ be  $[n,k]$ MDS code over $\F$ such that there
exists a uniform linear exact repair scheme for one erased
node, say $c_0,$ with bandwidth $n-1,$ i.e., the remaining 
codesymbols send one $\B$-subsymbol.
If $n<q^m$, then
\[
k<\left(1-\frac{1}{q}\right)n.
\]
\end{theorem}

\begin{proof}
By the dual characterization of linear repair schemes, there exist
codewords $h^{(1)},\ldots,h^{(m)}\in C^\perp$ such that condition \eqref{eq:gw:basic:1} and 
\eqref{eq:gw:basic:m} are satisfied.
Define the $\B$-linear subspace
\[
D=\operatorname{span}_{\B}\left\{h^{(1)},\ldots,h^{(m)}\right\}\subseteq C^\perp.
\]
Since $h_0^{(1)},\ldots,h_0^{(m)}$ are $\B$-linearly independent, the dual codewords
$h^{(1)},\ldots,h^{(m)}$ are also $\B$-linearly independent. Hence, $\dim_{\B}(D)=m.$
Consider the $\B$-linear map 
\begin{equation*}
    {\pi} \colon\left\{\begin{aligned}
        D & \longrightarrow \F \\
        h & \longmapsto h_0
    \end{aligned} \right.
\end{equation*}
Since for each $\ell\in [m]$ we have $\pi(h^{(\ell)})=h_0^{(\ell)},$ the dual codewords $h^{(1)},\ldots,h^{(m)}$ map directly on a 
$\B$-basis of $\F$, the map $\pi$ is an isomorphism.
Thus, for every $a\in\F$, there exists a unique $h_a\in D$ such that $(h_a)_0=a.$
%Since $D\subseteq C^\perp$, we have $(h_a)\in C^\perp$ for every $a\in\F$.
For every remaining codesymbol $j\in[n-1]$, define the $\B$-linear map
\begin{equation*}
    {L_j} \colon\left\{\begin{aligned}
        \F & \longrightarrow \F \\
        a & \longmapsto (h_a)_j.
    \end{aligned} \right.
\end{equation*}
Its image satisfies
$\im_\B(L_j)
\subseteq
\operatorname{span}_{\B}
\left\{
h_j^{(1)},\ldots,h_j^{(m)}
\right\}.$
Thus, by condition \eqref{eq:gw:basic:1}, we have $\dim_{\B}(\operatorname{Im}_\B(L_j))\leq1.$
Equivalently, $\operatorname{rk}_{\B}(L_j)\leq1.$
Since $\F$ has dimension $m$ over $\B$, the rank-nullity theorem gives
$\dimB(\ker(L_j))= \dimB(\F)-\rankB(L_j) \geq m-1.$
Hence,
$|\ker(L_j)|\geq q^{m-1}.$
Consequently, for every $j\neq 0$,
\begin{equation}\label{eq:first_count}
    \left|\left\{a\in\F\backslash \{0\}:L_j(a)\neq0\right\}\right|\leq q^m-1-(q^{m-1}-1)=q^{m-1}(q-1).
\end{equation}
Note that, since $C$ is a $[n,k]$ MDS code, it holds $d_H(C^\perp)=k+1.$
Fix $a\in\F\backslash\{0\},$ then by construction, $(h_a)_0=a\neq0.$
Thus, $h_a$ is a nonzero codeword of $C^\perp$, with at least $k+1$ not zero symbols.
Since the first coordinate is already nonzero, at least $k$ of the
remaining $n-1$ coordinates must be nonzero. Hence,
\begin{equation}\label{eq:second_count}
    \left|\left\{j\in[n-1]\setminus\{0\}:L_j(a)\neq0\right\}\right|\geq k
\end{equation}
for every $a\in\F\setminus\{0\}$.
We now count the set
\[
S=\left\{(a,j)\in\F\setminus\{0\}\times[n-1]:L_j(a)\neq0\right\}
\]
from both sides.
On the one hand, by \eqref{eq:second_count}, every $a\in\F\setminus\{0\}$ contributes
at least $k$ elements to $S$. Thus,
\begin{equation}\label{eq:third_count}
    |S|\geq k(q^m-1).
\end{equation}
On the other hand, by \eqref{eq:first_count}, each helper node $j\in[n-1]$ contributes at
most $q^{m-1}(q-1)$ elements to $S$. Since there are $n-1$ helper nodes, we get
\begin{equation}\label{eq:fourt_count}
    |S|\leq(n-1)q^{m-1}(q-1).
\end{equation}
Combining \eqref{eq:third_count} and \eqref{eq:fourt_count}, results into
\begin{equation}\label{fifth_count}
    k(q^m-1)\leq(n-1)q^{m-1}(q-1).
\end{equation}
Suppose, for contradiction, that $k\geq\left(1-1/q\right)n=(q-1)/q \cdot n.$
Then \eqref{fifth_count} gives $q^m\leq n.$
This contradicts the assumption $n<q^m.$
Therefore, $k<\left(1-1/q\right)n.$
\end{proof}

\section{Conclusion}\label{sec13}
In this work, we investigated the construction and algebraic properties of linear exact repair schemes 
through both theoretical and computational approaches.

We first revisited the explicit construction of Guruswami and Wootters and explored several ways of deriving 
additional LERSs. In particular, we considered Reed--Solomon codes of lower dimension, 
for which repair schemes can be obtained without requiring information from all remaining nodes. 
We also introduced a reverse transformation of check-polynomials for full-length Reed--Solomon codes, 
resulting an additional LERS. 
These results provide further insight into the algebraic structure of the Guruswami--Wootters construction 
and of LERSs that can be derived from it.

We then developed and implemented a systematic program for determining all uniform LERSs of bandwidth 
at most $n-1$ for any given linear code over $\F$. 
We applied this program to Reed--Solomon, Reed--Muller and Tamo--Barg codes, which  
revealed new LERSs for parameter regimes not covered by existing constructions. 
More precisely, for $\GRM(3,2)$ over $\mathbf{F}_{2^2}$, our program identified 491 distinct LERSs, despite the parameter 
condition required by the construction in \cite{CZ19} not being satisfied. 
Among these schemes, we found explicit LERSs with bandwidth $12$, allowing three remaining nodes to be omitted entirely in the 
repair process. 
Furthermore, we derived new LERSs for the Tamo--Barg codes $\TB(15,6,4),$ $\TB(15,5,4),$ and $\TB(8,4,3),$
with bandwidths $11,$ $11,$ and $5,$ respectively, which are, to the best of our knowledge, 
not covered by pervious constructions. 
Beyond these computational findings, our experiments motivated the formulation and proof of additional 
algebraic properties of LERSs.

Several directions for further research remain open. 
One direction is to 
improve the computational scalability of the proposed program.  
Although our hybrid strategy makes the systematic search more efficient, 
the computational complexity of the Gröbner basis approach remains a limiting factor for larger code parameters.
Exploiting additional algebraic constraints and code-specific properties could help reduce this complexity, resulting
in a larger feasible parameter range.
Another direction is to investigate whether the structural 
properties of the newly identified Reed--Muller and Tamo--Barg repair schemes can be used to derive 
more general algebraic constructions.

%%===================================================%%
%% For presentation purpose, we have included        %%
%% \bigskip command. Please ignore this.             %%
%%===================================================%%

\bibliography{sn-bibliography}% common bib file
%% if required, the content of .bbl file can be included here once bbl is generated
%%\input sn-article.bbl

\begin{appendices}

\section{Further Experimental Results for Reed--Solomon Codes}\label{Appendix:exper_results}

The experiments were performed on the Margaret high-performance computing cluster, 
which provides 512 CPU cores and $1024$ GiB of RAM.
Note that, in all experiments, the number of CPU cores is chosen to match the number of workers, 
such that each worker is assigned a dedicated CPU core. 
Recall, the program solves the algebraic system~\eqref{eq:grobner_basis_algebraic_system} 
and therefore computes all sets of dual codewords that 
induce a uniform LERS with bandwidth at most $n-1$.
As described in subsection~\ref{sec:hybrid_approach}, the algebraic 
system~\eqref{eq:grobner_basis_algebraic_system} can be simplified by reducing 
the number of unknowns. In our experiments, 
we considered two options: fixing $n-1$ unknowns or fixing $2(n-1)$ unknowns.
We also considered an additional simplification, controlled by the 
parameter $\ell$. More precisely, we considered the case in which the last $\ell$ columns (a.k.a. positions) are fixed to zero, 
which further reduces the number of unknowns and allows us to search 
for LERSs with bandwidth at most $n-1-\ell$.
The column $T$ (resp. $T_{\mathrm{res}}$) report 
the wall-clock running time for the completion of the program without (resp. with) the integrated resultant method. 
All running times are given in seconds. 
A dash indicates that the corresponding experiment was either not performed or did not terminate successfully.
The column ``LERS“ gives the number of LERSs identified by the program. Each distinct set of valid dual 
codewords is counted as one LERS. Finally, the ``Status“ column, if given, indicates whether the 
computation terminated successfully (reported with ``Completed“) or was terminated prematurely, 
for example due to resource limitations (reported with ``Terminated“). If the ``Status“ 
column is omitted then it can be assumed that all 
displayed experiments successfully completed.

Let $\F$ be a finite field of size $q^m$, and consider a Reed--Solomon code $\RS(\alpha,k)_\F$ of 
length $n$ with generator matrix $G$. The ordering of the columns of $G$ determines the ordering of the 
code coordinates and, equivalently, of the evaluation points. If not otherwise mentioned, throughout our experiments for Reed--Solomon codes, 
we use the evaluation points
$\alpha=(0,1,a,a^2,\ldots,a^{n-2}),$
where $a\in\F$ is a primitive element of $\F$. We assume that the first code coordinate, 
corresponding to the evaluation point $\alpha_0=0$, is erased and has to be repaired.

While conducting our experiments our strategy was as follows:  
Fixing the parameters $q,m$ and $k$, we first chose as the number of fixed 
unknowns/variables $n-1$ and, if the computation did not terminate 
 due to memory limitations, repeated the experiment with $2(n-1)$ fixed 
 unknowns.
If the system could not be solved with either $n-1$ or $2(n-1)$ fixed 
unknowns, we sometimes applied this additional simplification starting with $\ell=1$. 
Note, that this additional restriction changes the code under 
consideration: fixing the last $\ell$ positions to zero
corresponds to considering the punctured code at these positions, 
rather than the original code. Consequently, the resulting solutions do not, 
in general, include \textit{all} LERSs for the original code.

The results are presented in Table~\ref{table:RS:q2} for the binary case 
and in Table~\ref{table:RS:q3} for the non-binary case with $q=3$.

\begin{table}[h]
    \centering
    \setlength{\tabcolsep}{2.5pt}
    \begin{tabular}{ccccccccccc}
        \toprule
        $q$ & $m$ & $n$ & $k$ & $\ell$ &
         Workers &
         Fixed var. &
        $T_{\mathrm{res}}$ &
        $T$ &
        LERSs &
        Status \\
        \midrule

        2 & 2 & 3 & 1 & 0 & 256 & $n-1$ &2.17 & --  &6 & Completed \\
        2 & 2 & 4 & 1 & 0 & 20 & $n-1$ & 0.10 & 0.05 & 60 & Completed \\
        2 & 2 & 4 & 2 & 0 &  20 & $n-1$ &0.04 & 1.00 & 3 & Completed \\

        \midrule

        2 & 3 & 4 & 1 & 0 & 20 & $n-1 $& 0.11 & 0.07 &168 & Completed \\
        2 & 3 & 4 & 2 & 0 & 20 & $n-1$ & -- & 0.20  & 0 & Completed \\
        2 & 3 & 6 & 2 & 0 & 20 & $n-1$ & -- & 0.20 & 686 & Completed \\
        2 & 3 & 7 & 2 & 0 & 20 &$ n-1$ & 4026.38 & -- & 49168 & Completed \\
        2 & 3 & 7 & 3 & 0 & 20 & $n-1$ & -- & 0.09 & 28 & Completed \\
        2 & 3 & 8 & 2 & 0 & 256 & $2(n-1)$ & -- & 411544.28 & 2760296 & Completed \\
        2 & 3 & 8 & 2 & 2 & 20 & $n-1$ & -- & 0.19 & 686 & Completed \\
        2 & 3 & 8 & 3 & 0 &  20 & $n-1$ &-- & 8.47 & 3493 & Completed \\
        2 & 3 & 8 & 4 & 0 &  20 & $n-1$ &0.17 & 0.09 & 14 & Completed \\

        \midrule

        2 & 4 & 4 & 1 & 0 & 20 & $n-1$ & -- & 0.67 & 0 & Completed \\
        2 & 4 & 5 & 1 & 0 & 20 & $n-1$ & -- & 2.86 & 20160 & Completed \\
        2 & 4 & 5 & 2 & 0 &  20 & $n-1$ &-- & 0.32 & 0 & Completed \\
        2 & 4 & 5 & 3 & 0 & 20 &$ n-1$ & -- & 0.27 & 0 & Completed \\
        % 2 & 4 & 7 & 1 & 0 & 256 & $2(n-1)$ & -- & -- & -- & Terminated \\
        % 2 & 4 & 10 & 3 & 0 & 256 & $2(n-1)$ & -- & -- & -- & Terminated \\
        2 & 4 & 10 & $\leq 3$ & 0 & 256 & $2(n-1)$ & -- & -- & -- & Terminated \\
        2 & 4 & 10 & 4 & 0 & 256 & $2(n-1)$ & -- & 143.71 & 75 & Completed \\
        2 & 4 & 10 & 5 & 0 & 256 & $n-1$ & -- & 1.28 & 0 & Completed \\
        2 & 4 & 14 & 6 & 0 & 256 & $n-1$ & -- & -- & $\geq 147$ & Terminated \\
        2 & 4 & 14 & 7 & 0 & 256 & $n-1$ & -- & 35.66 & 0 & Completed \\
        % 2 & 4 & 15 & 3 & 0 & 256 & $2(n-1)$ & -- & -- & -- & Terminated \\
        2 & 4 & 15 & $\leq6$ & 0 & 256 & $2(n-1)$ & -- & -- & -- & Terminated \\
        2 & 4 & 15 & 7 & 0 & 256 & $2(n-1)$ & 3245.21 & 386 & 60 & Completed \\
        2 & 4 & 15 & 8 & 0 & 256 & $n-1$ &  -- & 19.79 & 0 & Completed \\
        2 & 4 & 15 & 13 & 0 & 256 & $n-1$ &  -- & 0.42 & 0 & Completed \\
        % 2 & 4 & 16 & 5 & 0 & 256 & $2(n-1)$ &  -- & -- & -- & Terminated \\
        2 & 4 & 16 & $\leq6$ & 0 & 256 & $2(n-1)$ &  -- & -- & -- & Terminated \\
        2 & 4 & 16 & 7 & 0 & 256 & $2(n-1)$ &  -- & -- & $\geq 59$ & Terminated \\
        2 & 4 & 16 & 8 & 0 & 256 & $2(n-1)$ & 413.12 & 1539.94 & 30 & Completed \\
        
        \midrule

        % 2 & 5 & 6  & 1  & 0 & 256 & $n-1$ & -- & -- & -- & Terminated \\
        2 & 5 & 6  & 1  & 1 & 256 & $n-1$ &-- & 0.31 & 0 & Completed \\
        2 & 5 & 6  & 2  & 0 & 20 & $n-1$ & -- & 1.19 & 0 & Completed \\
        2 & 5 & 6  & 3  & 0 & 20 & $n-1$ & -- & 0.38 & 0 & Completed \\
        % 2 & 5 & 14 & 5  & 0 & 64 & $n-1$ & -- & -- & -- & Terminated \\
        2 & 5 & 14 & $\leq 6$  & 0 & 64 & $n-1$ & -- & -- & -- & Terminated \\
        2 & 5 & 14 & 7  & 0 & 128 & $n-1$ & -- & 200.40 & 0 & Completed \\
        2 & 5 & 16 & 8  & 0 & 128 & $n-1$ & -- & 21370.25 & 0 & Completed \\
        2 & 5 & 18 & 9  & 0 & 64 & $n-1$ & -- & 13367 & 0 & Completed \\
        % 2 & 5 & 20 & 10 & 0 & 128 & $n-1$ & -- & -- & -- & Terminated \\
        % 2 & 5 & 28 & 14 & 0 & 256 & $n-1$ & -- & -- & -- & Terminated \\
        2 & 5 & 32 & $\leq 16$ & 0 & 256 & $n-1$ & -- & -- & -- & Terminated \\
        
        \bottomrule
    \end{tabular}
    \caption{Computational results for the enumeration of uniformly linear exact
    repair schemes with reduced bandwidth for Reed--Solomon codes over the binary field
    $\mathbb{F}$.}
    \label{table:RS:q2}
\end{table}

\begin{table}[h]
    \centering
    \setlength{\tabcolsep}{2.5pt}
    \begin{tabular}{ccccccccccc}
        \toprule
        $q$ & $m$ & $n$ & $k$ & $\ell$ &
        Workers &
        Fixed var. &
        $T_{\mathrm{res}}$ &
        $T$ &
        LERSs &
        Status \\
        \midrule
        3 & 2 & 4 & 1 & 0 & 256 & $n-1$ &-- &  & 432 & Completed \\

        3 & 2 & 5 & 2 & 0 & 256 & $n-1$ &-- &  & 168 & Completed \\

        3 & 2 & 6 & 1 & 1 & 256 & $n-1$ &-- & 93.52 & 14580 & Completed \\
        3 & 2 & 6 & 2 & 0 & 256 & $n-1$ & -- & 53.98 & 5868 & Completed \\
        3 & 2 & 6 & 3 & 0 & 256 & $n-1$ & -- & 0.68 & 68 & Completed \\
        3 & 2 & 6 & 4 & 0 & 256 & $n-1$ &-- & 0.70 & 0 & Completed \\

        3 & 2 & 7 & 4 & 0 & 256 & $n-1$ &-- &  & 20 & Completed \\

        3 & 2 & 8 & 1 & 3 & 256 & $n-1$ & -- & 92.94 & 14580 & Completed \\
        3 & 2 & 8 & 2 & 2 & 256 & $n-1$ & -- & 53.35 & 5868 & Completed \\
        3 & 2 & 8 & 3 & 1 & 256 & $n-1$ & -- & 9.51 & 2328 & Completed \\
        3 & 2 & 8 & 4 & 1 & 256 & $n-1$ & -- & 0.72 & 20 & Completed \\
        3 & 2 & 8 & 4 & 0 & 256 & $n-1$ &-- & 3.88 & 896 & Completed \\
        3 & 2 & 8 & 5 & 0 & 256 & $n-1$ &-- & 1.59 & 8 & Completed \\
        3 & 2 & 8 & 6 & 0 & 256 & $n-1$ &-- & 0.86 & 0 & Completed \\

        3 & 2 & 9 & 1 & 4 & 256 & $n-1$ & -- & 93.06 & 14580 & Completed \\
        3 & 2 & 9 & 2 & 3 & 256 & $n-1$ &-- & 53.45 & 5868 & Completed \\
        3 & 2 & 9 & 3 & 2 & 256 & $n-1$ &  -- & 9.15 & 2328 & Completed \\
        3 & 2 & 9 & 4 & 1 & 256 & $n-1$ & -- & 4.31 & 896 & Completed \\
        3 & 2 & 9 & 4 & 0 & 256 & $n-1$ & 30052.91 & -- & 31608 & Completed \\
        3 & 2 & 9 & 5 & 0 & 256 & $n-1$ & -- & 7.38 & 292 & Completed \\
        3 & 2 & 9 & 6 & 0 & 256 & $n-1$ & 4.79 & 4.22 & 4 & Completed \\

        \midrule

        3 & 3 & 4 & 1 & 0 & 256 & $n-1$ &-- & 0.53 & 1404 & Completed \\
        3 & 3 & 4 & 2 & 0 & 20 & $n-1$ & -- & 0.25 & 0 & Completed \\

        % 3 & 3 & 6 & 1 & 0 & 256 & $2(n-1)$ & -- & -- & -- & Terminated \\
        3 & 3 & 6 & 2 & 0 & 64 & $n-1$ & -- & 2177.99 & 7878 & Completed \\
        3 & 3 & 6 & 3 & 0 & 64 & $n-1$ & -- & 0.61 & 0 & Completed \\
        3 & 3 & 6 & 4 & 0 & 64 & $n-1$ & -- & 0.82 & 0 & Completed \\

        % 3 & 3 & 9 & 1 & 3 & 256 & $n-1$ & -- & -- & -- & Terminated \\
        % 3 & 3 & 9 & 2 & 2 & 256 & $n-1$ & -- & -- & -- & Terminated \\
        % 3 & 3 & 9 & 3 & 0 & 256 & $n-1$ & -- & -- & -- & Terminated \\
        3 & 3 & 9 & $\leq3$ & 0 & 256 & $n-1$ & -- & -- & -- & Terminated \\
        3 & 3 & 9 & 4 & 0 & 256 & $n-1$ & -- & -- & $\geq 654$ & Terminated \\
        3 & 3 & 9 & 4 & 1 & 256 & $n-1$ & -- & 5.18 & 0 & Completed \\
        3 & 3 & 9 & 5 & 0 & 128 & $n-1$ &  -- & 9.78 & 0 & Completed \\

        3 & 3 & 20 & 10 & 0 & 20 & $n-1$ & -- & 1665 & $\geq 1$ & Terminated \\

        % 3 & 3 & 27 & 18 & 0 & 256 & $n-1$ &-- & -- & -- & Terminated \\
        3 & 3 & 27 & 15 & 1 & 256 & $n-1$ & -- & 20150 & 0 & Completed \\
        % 3 & 3 & 27 & 14 & 3 & 256 & $n-1$ & -- & -- & -- & Terminated \\

        \bottomrule
    \end{tabular}
    \caption{Computational results for the enumeration of uniformly linear exact
    repair schemes with reduced bandwidth for Reed--Solomon codes over the field
    $\mathbb{F}$ with $q=3$.}
    \label{table:RS:q3}
\end{table}

%%=============================================%%
%% For submissions to Nature Portfolio Journals %%
%% please use the heading ``Extended Data''.   %%
%%=============================================%%

%%=============================================================%%
%% Sample for another appendix section			       %%
%%=============================================================%%

%% \section{Example of another appendix section}\label{secA2}%
%% Appendices may be used for helpful, supporting or essential material that would otherwise 
%% clutter, break up or be distracting to the text. Appendices can consist of sections, figures, 
%% tables and equations etc.

\end{appendices}

%%===========================================================================================%%
%% If you are submitting to one of the Nature Portfolio journals, using the eJP submission   %%
%% system, please include the references within the manuscript file itself. You may do this  %%
%% by copying the reference list from your .bbl file, paste it into the main manuscript .tex %%
%% file, and delete the associated \verb+\bibliography+ commands.                            %%
%%===========================================================================================%%

\end{document}